\documentclass[a4paper,11pt]{amsart}

\usepackage{color}
\usepackage{amssymb}

\usepackage{cite}

\newcommand{\bM}{{\overline M}}
 \newcommand{\delg}{{\gdelta}}

\newcommand{\zW}{{\mathring W}}
\newcommand{\Pbd}{P_{\bdelta }}

\newcommand{\gn}{{g_n}}
\newcommand{\hgn}{{\hat g_n}}

\newcommand{\Mext}{M_{\rm \scriptsize \mbox{ext}}}

\newcommand{\sqrtV}{\zN}

\newcommand{\bdelta}{{{} b_\delta}}
\newcommand{\gdelta}{{{}  g_\delta}}

\newcommand{\bmanN}{N}

\newcommand{\Vm}{V_{(\mu)}}
\newcommand{\Vmd}{V_{\delta,(\mu)}}
\newcommand{\Vnd}{V_{\delta,(\nu)}}

\newcommand{\dmud}{d\mu_{\bdelta }}
\newcommand{\dmutgd}{d\mu_{\widetilde{g}_{\delta,p_{(\mu)}}}}

\newcommand{\zN}{\mathring N}

\newcommand{\zD}{{\mathring D}}

\newcommand{\ourU}{\mathbb U}

\newcommand{\zh}
{\,\,{\mathring{\!\! h}}}

\newcommand{\KDE}{K_{\bdelta }}

\newcommand{\Esym}{E}

\newcommand{\bpzm}{b_{\pzm}}
\newcommand{\bdpzm}{b_{\delta,\pzm}}
\newcommand{\bdpm}{b_{\delta,\pgm}}
\newcommand{\pzm}{p{}^0_{(\mu)}}
\newcommand{\pgm}{{p_{(\mu)}}}
\newcommand{\pzgm}{{p_{(0)}}}
\newcommand{\qgm}{{q_{(\mu)}}}

\renewcommand{\proof}{{\noindent \sc Proof:}\ }

\definecolor{bluem}{rgb}{0,0,0.5}

\definecolor{mycolor}{cmyk}{0.5,0.1,0.5,0}
\definecolor{michel}{rgb}{0.5,0.9,0.9}

\definecolor{turquoise}{rgb}{0.25,0.8,0.7}
\definecolor{bluem}{rgb}{0,0,0.5}

\definecolor{MDB}{rgb}{0,0.08,0.45}
\definecolor{MyDarkBlue}{rgb}{0,0.08,0.45}

\definecolor{MLM}{cmyk}{0.1,0.8,0,0.1}
\definecolor{MyLightMagenta}{cmyk}{0.1,0.8,0,0.1}

\definecolor{HP}{rgb}{1,0.09,0.58}

\newcommand{\ptcx}[1]{}

\newcommand{\Fd} {f_{\gdelta }}%

\newcommand{\hpz}{z'}

\newcommand{\Ik}{I_k}

\newcommand{\mcO}{\mycal O}
\newcommand{\mcU}{\mycal U}
\newcommand{\hmcU}{\,\widehat{\! \mcU}}

\newcommand{\gd}{{\gdelta }}

\newcommand{\Hkd}{{H^k_\delta}}

\newcommand{\loc}{{\textrm{loc}}}

\newcommand{\dte}
{\delta^{4}}
\newcommand{\dnte}
{\delta_n^{4}}

\newcommand{\hb}{{\hat b}}

\newcommand{\dn}{\delta _n}

\DeclareFontFamily{OT1}{rsfs}{}
\DeclareFontShape{OT1}{rsfs}{m}{n}{ <-7> rsfs5 <7-10> rsfs7 <10->
rsfs10}{} \DeclareMathAlphabet{\mathscr}{OT1}{rsfs}{m}{n}

\newcommand{\eq}[1]{\eqref{#1}}

\newcommand{\bel}[1]{\begin{equation}\label{#1}}
\newcommand{\bea}{\begin{eqnarray}}
\newcommand{\beaa}{\begin{eqnarray*}}
\newcommand{\bean}{\begin{eqnarray}\nonumber}
\newcommand{\beal}[1]{\begin{eqnarray}\label{#1}}
\newcommand{\beadl}[1]{\begin{deqarr}\label{#1}}
\newcommand{\eeadl}[1]{\arrlabel{#1}\end{deqarr}}
\newcommand{\eeal}[1]{\label{#1}\end{eqnarray}}
\newcommand{\eead}[1]{\end{deqarr}}
\newcommand{\eea}{\end{eqnarray}}
\newcommand{\eeaa}{\end{eqnarray*}}

\newcommand{\be}{\begin{equation}}
\newcommand{\ee}{\end{equation}}

\DeclareFontFamily{OT1}{rsfs}{}
\DeclareFontShape{OT1}{rsfs}{m}{n}{ <-7> rsfs5 <7-10> rsfs7 <10->
rsfs10}{} \DeclareMathAlphabet{\mycal}{OT1}{rsfs}{m}{n}

\newcounter{mnotecount}[section]

\newcommand{\N}{{\Bbb N}}

\newcommand{\rmnote}[1]{}

\newcommand{\Ric}{\operatorname{Ric}}

\def\mysavedown#1{\edef\mysubs{\mysubs#1}}
\def\mysaveup#1{\edef\mysups{\mysups#1}}
\def\mydown#1{{\mytensor}_{\vphantom{\mysubs}#1}}
\def\myup#1{{\mytensor}^{\vphantom{\mysups}#1}}
\def\tensor#1#2{
  #1
  \def\mytensor{\vphantom{#1}}
  \def\mysubs{\relax}
  \def\mysups{\relax}
  \let\down=\mysavedown
  \let\up=\mysaveup
  #2
  \let\down=\mydown
  \let\up=\myup
  #2
  }

\newcommand{\Ker}{\operatorname{Ker}}

\newcommand{\Tr}{\operatorname{Tr}}

\newcommand{\R}{\mathbb R}
\newcommand{\Sp}{\mathbb S}

\renewcommand{\setminus}{\smallsetminus}

\renewcommand{\to}{\rightarrow}

\renewcommand{\epsilon}{\varepsilon}
\renewcommand{\hat}{\widehat}

\def\crn#1#2{{\vcenter{\vbox{
        \hbox{\kern#2pt \vrule width.#2pt height#1pt
           }
          \hrule height.#2pt}}}}

\newcommand{\hyp}{M}

\renewcommand{\H}{\mathbb H}

\renewcommand{\hbar}{{\overline h}}

\newcommand{\pre}[2]{{{\vphantom{#2}}^{#1}}\kern-.2ex{#2}}

\theoremstyle{plain}
\newtheorem{theorem}{\sc Theorem}[section]
\newtheorem{Theorem}[theorem]{\sc Theorem}

\newtheorem{Lemma}[theorem] {\sc Lemma}
\newtheorem{proposition}[theorem]{\sc Proposition}

\newtheorem{Corollary}[theorem] {\sc Corollary}

\theoremstyle{definition}
\newtheorem{definition}[theorem]{Definition}

\newtheorem{Remark}[theorem]{\sc  Remark\rm}

\numberwithin{equation}{section}

\date{\today}

\begin{document}

\title[Gluing constructions for negative scalar curvature] {Gluing constructions for asymptotically hyperbolic
manifolds with constant scalar curvature}
\author[P.T. Chru\'sciel]{Piotr T.~Chru\'sciel} \address{Piotr
T.~Chru\'sciel, LMPT, F\'ed\'eration Denis Poisson,
Tours; Mathematical Institute and Hertford College, Oxford.}
\email{chrusciel@maths.ox.ac.uk} \urladdr{
http://www.phys.univ-tours.fr$/\sim${piotr}} \author[E.
Delay]{Erwann Delay} \address{Erwann Delay, Laboratoire d'analyse
non lin\'eaire et g\'eom\'etrie, Facult\'e des Sciences, 33 rue
Louis Pasteur, 84000 Avignon, France}
\email{Erwann.Delay@univ-avignon.fr}
\urladdr{http://www.math.univ-avignon.fr/Delay}

\begin{abstract} We show that asymptotically hyperbolic
initial data satisfying smallness conditions in dimensions $n\ge 3$,
or fast decay conditions in $n\ge 5$,  or a genericity condition in
$n\ge 9$, can be deformed, by a deformation which is supported
arbitrarily far in the asymptotic region, to ones which are exactly
Kottler (``Schwarzschild- adS") in the asymptotic region.
\end{abstract}

\maketitle

\tableofcontents
\section{Introduction}\label{section:intro}

One of the key problems in mathematical general relativity is
the understanding of the space of solutions of the vacuum
constraint equations. In this context an important gluing
method has been introduced by Corvino and
Schoen~\cite{Corvino,CorvinoSchoen2} for vacuum data with
vanishing cosmological constant. The object of this paper is to
present related gluing results   when the cosmological constant
$\Lambda$ is negative. The question we address is the
possibility of deforming an asymptotically hyperbolic
Riemannian manifold of constant scalar curvature, and hence a
time-symmetric vacuum initial data set, to one with a Kottler
metric (sometimes known as Schwarzschild -- anti de Sitter
metric) outside of a compact set. We establish deformation or
extension theorems in dimensions $n\ge 3$ under a smallness
condition for metrics sufficiently close to (generalized)
Kottler metrics, or under smallness and parity conditions for
metrics close to a standard hyperbolic metric, or assuming a
rapid decay condition in dimensions $n\ge 5$.

More precisely, we consider $n$-dimensional manifolds containing
asymptotic ends
\bel{Mext}
 \Mext:=(r_0,\infty)\times N
  \;,
\ee
where $N$ is a compact manifold. We are  interested in constant
scalar curvature metrics which asymptote, as $r$ goes to
infinity, to a background metric $b$ of the form%
\footnote{The constant $k$ in \eq{bmetinr}-\eq{bmetinr2} is of
course unrelated to the order of differentiability $k$ used
elsewhere, we hope that this will not confuse the reader.}
\bel{bmetinr}
 b = \frac {dr^2} {r^2 + k }
 + r^2 \hat b\;,
\ee
where $k\in \{0,\pm 1\}$, and where $\hat b$ is a
($r$--independent) metric on $N$ satisfying $\Ric(\hat b) =
k(n-2)\hat b$. A family of examples is provided by the
(generalized) Kottler metrics,
\bel{bmetinr2}
 b_m = \frac {dr^2} {r^2 + k  - \frac {2m}{r^{n-2}}}
 + r^2 \hat b\;.
\ee
Note that $b_0=b$, with $b$ as in \eq{bmetinr}.

 For the purpose of the next theorem define the manifold
$M$ to be
$$
 M= (r_0,r_2]\times N \;,
$$
and suppose that $g$ is a constant negative scalar curvature metric
on $M$ close to $b$, or to $b_m$. There are two natural questions:

First, choose $r_1$ satisfying $r_0<r_1<r_2$, can one deform $g$,
keeping the scalar curvature fixed, so that the resulting metric
coincides with $g$ on $(r_0,r_1]\times N$, and with $b_m$, for some
$m$, near $\{r_2\} \times N$? In this case we set
$M'=(r_0,r_1]\times N$, $M''=[r_2,\infty)\times N$, and we  refer to
this case as the \emph{deformation problem}.

Next, let $r_3>r_2$, can one extend $g$ to a new metric of
constant scalar curvature on $(r_0,\infty)\times N$ so that the
extended metric coincides with $b_m$, for some $m$, on
$[r_3,\infty)\times N$? In this case we set $M'=M$,
$M''=[r_3,\infty)\times N$, and we refer to this case as the
\emph{extension problem}.   It is shown
in~\cite[Section~8.6]{ChDelay} how to reduce this problem to
the deformation one.

Our aim here is to show that those problems can always be
solved when $g$ is sufficiently close to $b$, except perhaps
when $(N,\hat b)$ is a round sphere and $m=0$, in which case we
need to impose a restrictive condition: For $(r,q)\in M$ let
$\psi(r,q)=(r,\phi(q))$, where $\phi$ is the antipodal map of
the sphere. A metric $g$ on $M$ will be said to be
parity-symmetric if $\psi^* g = g$. At the end of
Section~\ref{A7} we prove:

\begin{Theorem}\label{TM1gN}
Let $n\geq 3$, $\N  \ni \ell>\lfloor \frac{n}{2}\rfloor + 4$,
$\lambda\in (0,1)$, $m\in \R$. If $(N,\hat b)$ is   a round sphere
\emph{and} $m=0$, we suppose moreover that $g$ is parity-symmetric.
There exists $\epsilon>0$ such that if $\|g-b_m\|_{C^{\ell ,\lambda
}(M)}<\epsilon$, then there exists on $\Mext$ a $C^{\ell ,\lambda }$
metric of constant negative scalar curvature which   coincides with
$g$ on $  M'$, and which is a Kottler metric on $M''$. If $g$ is
smooth, then so is the solution of the deformation problem.
\end{Theorem}

 We emphasize that $g$ and $b_m$ are only required to be
close to each other on an ``annulus" as above, and in fact
$b_m$ is not even defined throughout the original manifold.

We can also prove a result without smallness assumptions which,
however, excludes dimensions three and four. Moreover  the
decay rates  are undesirably restrictive in dimensions five,
six and seven; they are satisfactory, but not as weak as one
would wish, in higher dimensions.

Let $g$ be an asymptotically hyperbolic metric as defined in
Section~\ref{sec:def}, and let $\pzm$ be the momentum vector of
$g$, obtained by passing to the limit as $r$ goes to infinity
of the integral \eq{pint} below over submanifolds $r=\mbox{\rm
const}$. Let $b_{p_{(\mu)}}$ denote a (generalized, boosted)
Kottler metric with momentum vector $\pgm$. Suppose that
\bel{ineq2i} n\ge 5\;,\quad
 \alpha>\left\{ \begin{array}{cc}
                                        8\;, & n=5,6,7, \\
                          \frac{8+n }2\;, &
                             n \ge     8.
                     \end{array}
 \right.
\ee
(This can be compared with the   conditions $\alpha>n/2$, $n\ge
3$, needed for $\pzm$ to be  well-defined, or $\alpha=n$, which
holds for  Kottler metrics.) For $\alpha\le n$ we assume that
$g$ has the following asymptotic behaviour
\bel{condg1i} |g-b|_b+|\zD( g-b)|_b+\ldots
+|\zD^{(k+2)}(g-b)|_b=O(\rho^\alpha)
 \;,
\ee
where $\zD$ is the covariant derivative operator of $b$.
  For $\alpha>n$,
if $(N,\hat b)$ is a round sphere, we assume that the momentum
vector $\pzm$ of $g$ is
\emph{timelike},%
\footnote{Both past and future pointing $\pzm$ are allowed.}
so that an associated (perhaps boosted) Kottler metric $\bpzm$
exists. Whether or not $(N,\hat b)$ is a round sphere,  for
$\alpha>n$ instead of \eq{condg1i} we suppose that
\bel{condg2i} |g-\bpzm|_b+|\zD( g-\bpzm)|_b+\ldots
+|\zD^{(k+2)}(g-\bpzm)|_b=O(\rho^\alpha)
 \;;
\ee
in fact, \eq{condg2i} is equivalent to \eq{condg1i} if $\alpha \le
n$.

Letting $M_\delta$ be as in \eq{Mdeltadef}, and $A_{\delta,4\delta}$
as in \eq{Andef},  in Section~\ref{Sgcma} we prove:

\begin{theorem}
 \label{Tm2intro}
Let $n\geq 5$, $\N  \ni \ell>\lfloor \frac{n}{2}\rfloor + 4$,
$\lambda\in (0,1)$, and let $\alpha>0$ satisfy \eq{ineq2i}. Let
$g$ be a $C^{\ell,\lambda}$ asymptotically hyperbolic metric
with constant negative scalar curvature satisfying \eq{condg2i}
with $k=\ell-4$. We furthermore assume that \eq{condg2i} holds
with $k=\ell-2$ and $\alpha=0$, and that the energy-momentum
vector is timelike if $(N,\hat b)$ is a round sphere.
There exists $\delta_0>0$ such that for all
$0<\delta\leq\delta_0$ the metric $g$ can be deformed across an
annulus $A_{\delta,4\delta}$ to a constant scalar curvature
metric, of $C^{\ell,\lambda} $ differentiability class, which
coincides with $g$ on $M \backslash M_{4\delta}$ and with a
Kottler metric $b_{p_{(\mu)}}$ on $M_{\delta}$.
 The solution is smooth if $g$ is.
\end{theorem}

A key role in our analysis is played by the kernel of the
operator $P^*_g$ given by \eq{Pstar} below; it is known that
this kernel is trivial for any open subset of $M$ for generic
metrics~\cite{ChBeignokids}. Deforming first the metric as in
Section~\ref{sbdb}, and applying Theorem~\ref{Tm2intro} to the
new metric  one concludes:

\begin{Corollary}
Let $n\ge 9$. Under the remaining hypotheses of
Theorem~\ref{Tm2intro}, suppose instead of \eq{ineq2i} that
$\alpha>n/2$. If there are no neighbourhoods of the conformal
boundary at infinity on which $P^*_g$ has a kernel, then the
conclusions of Theorem~\ref{Tm2intro} hold.
\end{Corollary}

 It might be helpful to the reader to recall briefly the
Corvino-Schoen method, as adapted to our setting. We work on an
end $(r_0,+\infty)\times N$, where we have a metric $g$
asymptotic to a background metric $b$. We also have a
$d$-parameter family of references metrics $b_p$, all
asymptotic to $b$, all having the same, constant scalar
curvature. The gluing is performed on an  annulus
$A_R=\{R<r<4R\}$,  with $R\gg1$,
in four steps:\\

Step 1): Do a scaling in order to work on a fixed annulus $A_1$,\\

Step 2): Establish a weighted estimate of the form
$|P^*u|_{L^2}\geq c|u|_{H^2}$, where $P^*$ is the adjoint of
the linearized scalar curvature operator $P$ and $u$ is
orthogonal to the $d$-dimensional kernel $K$ of $P^*$. The
constant $c$ has to be uniform in the family of metrics under
consideration, with controlled dependence upon $R$. In fact, in
previous applications $c$ was $R$--independent.\\

Step 3): By step 2), and  up to weighting functions, the
operator $L=PP^*$ is an isomorphism modulo projections onto
$K^\bot$. By the inverse function theorem, for $R\gg1$, the
gluing of $g$ with {\em any} $b_p$ can be done modulo
 weighted $L^2$-projection onto
$K^\bot$.\\

Step 4): Estimate the projection onto $K$ and show that you can
adjust the parameter $p$ to obtain a solution.\\

So, the overall strategy is the same as
in~\cite{Corvino,CorvinoSchoen2,ChDelay}. However, in our case
essential new difficulties arise:  the scaling transformation
in the asymptotically flat case leads to a family of uniformly
equivalent operators on a fixed annulus, while this is not the
case anymore for negative $\Lambda$. To handle this we prove a
sharp estimate on the family of operators which arise in our
context; unfortunately the estimate degenerates as the gluing
annuli recede to infinity, as  the sharp constant $c$ in step
2) above goes to zero. This results in the undesirable
restrictions described above. A possible approach to improve
this state of affairs could be to devise a method which, first,
deforms any asymptotically hyperbolic metric to one for which
our Theorem~\ref{Tm2intro} (or some variation thereof) applies.
Alternatively, a completely different method of approaching the
problem is needed.

Our work has been largely motivated
by~\cite{AnderssonGallowayCai}, to remove the sign condition on
the mass aspect function imposed there. Our deformation
produces a metric with a constant mass aspect without a priori
assuming such a sign in dimensions larger than eight; our
result is, however,  irrelevant for the main result
in~\cite{AnderssonGallowayCai}, which has only been proved so
far for $n\le 7$.

\section{Definitions, notations and conventions}\label{sec:def}

Let $\overline{M}$ be a smooth, compact $n$-dimensional
manifold with boundary $\partial {M}$. Let
$M:=\overline{M}\backslash\partial{M}$, a non-compact manifold
without boundary. In our context the boundary $\partial {M}$
will play the role of a \emph{conformal boundary at infinity}
of $M$.  We will choose a defining function $\rho$ for
$\partial M$, that is a non-negative smooth function on
$\overline{M}$, vanishing precisely on $\partial M$, with
$d\rho$ never vanishing  there.

We will work near the infinity of $M$, so it is convenient to
define, for small $\epsilon>0$, the manifold
\bel{Mdeltadef}
 M_\epsilon=\{x\in M, \rho(x)<\epsilon\}.
\ee
We also define for small $\epsilon>\delta>0$, the ``annulus"
\bel{Andef}
 A_{\delta,\epsilon}:=M_\epsilon\backslash \overline{M_\delta}.
\ee
We continue by defining a class of background metrics of interest.
For $k$ equal to $-1$, $0$ or $1$, let $\hat{b}$ be a metric on
$\partial M$ satisfying Ric$(\hat{b})=k(n-2)\hat{b}$. For $\rho_0$
such that $1-k(\frac\rho2)^2$ has no zeros on $(0,\rho_0)$, consider
the metric%
\bel{methypxx}
 b=\rho^{-2}\left(d\rho^2+\frac{(4-k \rho ^2)^2}{16}\hat{b}\right)=:\rho^{-2}\overline{b}
\ee
defined on $(0,\rho_0)\times \partial M$. Then $b$ is Einstein,
$\Ric(b)=-(n-1)b$, in particular it  has constant scalar curvature
$R(b)=-n(n-1)$, and in fact provides initial data for a static
solution of the vacuum Einstein equations with a negative
cosmological constant. These are of course identical to \eq{bmetinr}
(use $r=\rho^{-1}[1-k(\frac\rho2)^2]$). The basic example of such a
background is the standard hyperbolic metric. In that case $M$ is
the unit ball of $\R^n$, with
\bel{hyprep} b=\omega^{-2}\delta
 \;,
\ee
$\delta$ is the Euclidean metric,
$\omega(x)=\frac{1}{2}(1-|x|_\delta^2)$.

A metric $g$ will be called \emph{asymptotically hyperbolic} if $g$
tends to a background metric as in \eq{methypxx} when approaching
$\partial M$. The precise decay rates will be indicated whenever
needed. The terminology is motivated by the fact that the sectional
curvatures of $g$ tend to $-1$ as $\rho$ approaches zero; cf., e.g.,
\cite{Mazzeo:hodge}. One should, however, keep in mind that $b$ does
not necessarily have constant sectional curvature in space-time
dimension other than four. Moreover, metrics which asymptote to
hyperbolic metrics in cuspidal ends do not necessarily belong to our
class.

An important class of asymptotically hyperbolic metrics is
given by the \emph{(generalized) Kottler
metrics}~\cite{Kottler} (compare~\cite{Birmingham}) as given by
\eq{bmetinr2}. In the coordinate system of \eq{methypxx} they
read
\begin{eqnarray}
 \label{bKottler}
 b_m&=&\rho^{-2}\Big\{\left[1-2m\rho^n\left(1-k(\frac\rho2)^2\right)^{2-n}\left(1+k(\frac\rho2)^{2}\right)^{-2}\right]^{-1}d\rho^2
  \\
  \nonumber
  &&
  \hspace{6.5cm}+(1-k(\frac\rho2)^2)^2\hat{b}\Big\}
   \\&=&\rho^{-2}\left[\overline{b}+2m\rho^n(1+
     {O(\rho^{2}))}d\rho^2\right]
   \;,
 \nonumber
\end{eqnarray}
where, as before, $\hat b$ is a fixed metric on the boundary at
infinity $\bmanN$ satisfying Ric$(\hat{b})=k(n-2)\hat{b}$. Those
metrics satisfy $R(b_m)=-n(n-1)$, and again provide initial data for
static Einstein metrics.

If $\hat b$ is \emph{not} the round metric on a sphere, the
only energy-like Hamiltonian invariant of $b_m$ is $m$, see
e.g.~\cite{ChBamberg} and references therein. Otherwise $b$ is
the standard hyperbolic metric, and the energy-momentum vector
of $b_m$, say $\pgm$ (as defined in~\cite{ChHerzlich}
or~\cite{Wang},  see   \eq{decaygen} with
$r\rightarrow+\infty$)), is proportional to $(m,\vec 0)$. Under
isometries of hyperbolic space, $\pgm$ transforms as a Lorentz
vector, and a metric with any timelike $\pgm$ can be obtained
by applying such an isometry to some $b_m$.

In this way we generate a family of metrics with any
\emph{timelike} $\pgm$, as needed for the Brouwer fixed point
argument when compensating for the cokernel below. (On the
other hand we are not aware of existence of such metrics with
non-timelike non-zero $\pgm$, whence the restriction of
timelikeness in our results when $(N,\hat b)$ is a round
sphere.) We denote by $b_\pgm$ the resulting metrics, and we
will refer to them as Kottler metrics, or boosted Kottler
metrics when ambiguities are likely to occur.

Recall that the linearized scalar curvature operator $P=P_g$ is
$$
P_g h := DR(g)h=-\nabla^k\nabla_k(tr_g
\;h)+\nabla^k\nabla^lh_{kl}-R^{kl}h_{kl}
 \;,
$$
so that its $L^2$ formal adjoint reads
\bel{Pstar}
\begin{array}{lllll}
P^*_gf&=&[DR(g)]^*f&=&-\nabla^k\nabla_k fg+\nabla\nabla
f-f\;Ric(g)
 \;.
\end{array}
\ee
(We use the summation convention, indices are lowered with $g_{ij}$
and raised with  its inverse $g^{ij}$.) We note that
$$
\Tr P^*_g f=(n-1)\nabla^*\nabla f-R f
 \;.
$$

Let
\bel{bdelmet}
 \bdelta:= z^{-2}\Big(dz^2 + \delta^{-2} \hat b(\delta z)\Big)
\ee
be a hyperbolic metric scaled up in $\rho$ from $A_{\delta,4\delta}$
to
\bel{Adef}
 A\equiv A_{(1,4)}:= (1,4)\times\partial M
 \;.
\ee
We have $Ric(b_\delta)=-(n-1)b_\delta$, thus
\bel{Pdeleq}
 \Pbd ^* u:= \nabla \nabla u +\Big( (n-1) u- \nabla^k \nabla_ku\Big) \bdelta
 \;,
\ee
where  $\nabla$ is associated with the metric $\bdelta $.

It is well known that the kernel of $P^*$ has dimension at most
$n+1$, see~\cite{Corvino} for instance. For the hyperbolic
metric $b$ on the unit $n$-dimensional ball $B^{n}(1)\subset
\R^n$, in the representation \eq{hyprep}, the kernel of $P^*_b$
is spanned by the following functions,

which are the restrictions to the hyperboloid $\H^n$ of the
coordinates functions in Minkowski $\R^{n,1}$:
\beal{KID1} {V_{(0)}}&:=&{\frac{1+|x|^2}{1-|x|^2}}=
\rho^{-1}\left(1+ \left(\frac \rho 2 \right)^2\right)
 \;,
 \\ 
{V_{(k)}}&:=&-{\frac{2x^k}{1-|x|^2}}= -\rho^{-1}\left(1-
\left(\frac \rho 2 \right)^2\right)\frac {x^k}{|x|}\;,
 \eeal{KID2}
with $\rho= 2(1-|x|)/(1+|x|)$. We can also rewrite \eq{hyprep} as
\bel{hbmet}
 b= \rho^{-2}(d\rho^2 + \hat b (\rho))\;,
\ee
 {with} $
 \hat b(\rho)= \left(1-\left(\frac\rho 2\right)^2\right)^2
 \hat b(0)$, where $\hat b(0)$ is the round unit metric on $S^n$. Setting
$\rho=\delta z$, defining
\bel{Vdm}
V_{\delta,(\mu)}(z,\theta)=V_{(\mu)}(\delta z,\theta)
 \;,
\ee
and letting $\delta$ tend  to zero,  the functions $\delta
V_{\delta,(\mu)}$  tend to
\beal{KID3}
 {u_{(0)}}:=z^{-1} \;, \qquad {u_{(k)}}:= -z^{-1} \frac
 {x^k}{|x|}
  \;.
\eea

For  nonspherical boundary metrics $\hat b\equiv \hb(0)$ we
still write $\{V_{(\mu)}\}$ for any basis of $\Ker P^*_b$, and
then $V_{\delta,(\mu)}$ is defined by \eq{Vdm}. By hypothesis
the scalar curvature of   $\hb$ is constant, so that we can
invoke a theorem of Obata~\cite{Obata} (see also~\cite{Kuehnel}
Theorem 24) to conclude that the only Riemannian manifold
$(\hat{M},\hat{b})$ of dimension $n-1$ with non-constant
solutions $v$ to the equation $DDv+\frac{ D^*Dv}{n-1}\hat{b}=0$
is a  round sphere. For metrics of the form \eq{methypxx} with
$\hat b$ different from a round sphere, this implies  (compare
appendix \ref{ASc}) that $\dim \Ker P^*= 1$, and
$V_{(0)}=V_{(0)}(\rho)$, with the $u_{(\mu)}$'s proportional to
$u_{(0)}=z^{-1}$ (compare \eq{ww3} and \eq{ww4}).

\begin{definition}
 \label{D1}
Let $k\in \N$, $C,\sigma\ge 0$. Let $b$ be a  of the form
\eq{methypxx}, with $\hat b$ an Einstein metric on $\partial M$ with
scalar curvature $(n-1)(n-2)\kappa$, $\kappa\in \{0,\pm 1\}$, and
with $\rho\in (0,2\rho_0]$. We will say that $g$ is
$(C,k,\sigma)$\emph{-asymptotically hyperbolic} if  we have
\bel{gestimb}
  |g-b|_b+|\nabla g|_b+...+|\nabla^{(k)}g|_b\leq
 C\rho^\sigma\;,
\ee
where the norm and covariant derivatives are defined by $b$. For
$\alpha\in (0,1)$ we will say that $g$ is
$(C,k+\alpha,\sigma)$\emph{-asymptotically hyperbolic} if the
derivatives of order $k$ of $g-b$ further satisfy a weighted
H\"older condition of order $\alpha$, as in~\cite{Lee:fredholm}.
\end{definition}

Let $g$ be a Riemannian metric on $M$, recall that $(M,g)$ is {\it
conformally compact} if there exists on $\overline{M}$ a smooth
defining function $\rho$ for $\partial M$ (that is $\rho\in
C^\infty(\overline{M})$, $\rho>0$ on $M$, $\rho=0$ on $\partial {M}$
and $d\rho$ nowhere vanishing on $\partial M$; the symbol $\rho$
will  be used throughout this work to denote such a function) such
that $\overline{g}:=\rho^{2}g$ is a  Riemannian metric on
$\overline{M}$, we will denote by $\hat{g}$ the metric induced on
$\partial M$. The background metrics $b$ considered above are
conformally compact in this sense.

It is well know that, near infinity, for any sufficiently
differentiable conformally compact metric $g$ we may choose the
defining function $\rho$ to be the $\overline{g}$-distance to the
boundary. Thus, if $\epsilon$ is small enough, $M_\epsilon$ can be
identified with $(0,\epsilon)\times\partial M$ equipped with the
metric
\bel{metrichgold}
 g=\rho^{-2}(d\rho^2+\hat{g}(\rho))=\rho^{-2}(d\rho^2+\hat{g}_{AB}(\rho)d\theta^Ad\theta^B),
\ee
where $\{\hat{g}(\rho)\}_{\rho\in(0,\epsilon)}$ is a family of
smooth, uniformly equivalent, metrics on $\partial M$, with
$\hat{g}(0)=\hat{g}$. However, the introduction of this system
of coordinates might lead to a loss of up to two derivatives of
the metric. This can be circumvented for
$(C,k,\sigma)$-asymptotically hyperbolic metrics  by
introducing a coordinate system as
in~\cite[Appendix~B]{AndChDiss} in which $g$ takes the form
\bel{metrichgnew}
 g= \rho^{-2}\left((1+O(\rho^{k+\sigma}))d\rho^2+\hat{g}_{AB}(\rho)d\theta^Ad\theta^B
  +O(\rho^{k+\sigma})_Ad\rho d\theta^A\right),
\ee
with all metric coefficients of original differentiability class.

If  $g$ is $(C,k,\sigma)$-asymptotically hyperbolic with $k\ge 2$
and $\sigma>0$,  we have
\bel{Pdelegg1}
 P_{g}^* u:=
 \nabla \nabla u -\nabla^k \nabla_ku\, g-
 u\Ric(g)
 =\nabla \nabla u +\Big( (n-1) u- \nabla^k \nabla_ku\Big)
 g+ O(\rho^\sigma)u
 \;,
\ee where the covariant derivatives are related to $g$, and the
$O(\rho^\sigma)$ term is bounded (in $b$-norm) together with its
$b$-derivatives up to order $k-2$, by  $\rho^\sigma$ times a
constant depending on $C$ and $k$.

\section{A uniform estimate for $P^*$}\label{unifP}

Let $y$ be the function on $A$ defined by:
\bel{ydef}
\begin{array}{cccc}
y:&(1,4)\times\partial M&\longrightarrow&\R\;,\\
&(z,\theta)&\mapsto&\frac{4}{3}(1-\frac{z}{4})({z}{}-1)
 \;.
\end{array}
\ee
We claim that:

\begin{proposition}
 \label{Pcontr}%
 Let $c_0,\sigma>0$ and $s\ge 0$. There exist   constants $c_1=c_1(n,s,c_0,\sigma)>0$ and
$\delta_0=\delta_0(n,s,c_0,\sigma)>0$ such that for all
$(c_0,4,\sigma)$-asymptotically hyperbolic metrics  $g$, for all
$0<\delta\leq\delta_0$, and for all $u $ satisfying
\bel{orthoc}
 \forall\ \mu=0,\ldots,k \qquad \int_A e^{-s/y}u u_{(\mu)}d\mu_{\bdelta }=0
  \;,
\ee
where $k=n$ if $b$ is the standard hyperbolic metric, and $k=0$
otherwise, we have
\bel{wewant} \int_{A} e^{-s/y}y^8|P_{\gdelta }^* u|^2_{\gdelta
}d\mu_{\gdelta }\geq
{c_1}{\dte}
 \int_{A}
e^{-s/y}(y^8|\nabla \nabla u|^2_{\gdelta }+y^4|\nabla u|^2_{\gdelta
}+u^2)d\mu_{\gdelta }
 \;,
\ee
%
provided that the right-hand-side is finite. Similarly \eq{wewant}
holds (with perhaps a different constant $c_1$) if
\bel{orthocv}
 \forall\ \mu=0,\ldots,k \qquad \int_A e^{-s/y}u V_{\delta,(\mu)}d\mu_{\gdelta }=0
 \;,
\ee
or if in \eq{orthoc} the measure $z^{-n}dz\, d\mu_{\hat b(0)}$ is
used.
\end{proposition}
\begin{Remark}
\label{Rdelta} There is little doubt that the result remains valid
for $(c_0,2,\sigma)$ asymptotically hyperbolic metrics, or  for
those conformally compact metrics which are $C^2$ up-to-boundary
after the conformal rescaling, by using coordinates as in
\eq{metrichgnew}. For simplicity of calculations we assume
\eq{metrichgold}, since our main gluing results require
$(c_0,4,\sigma)$ asymptotically hyperbolic metrics anyway.
\end{Remark}

\begin{Remark}
\label{Pcontrsharp} The power of $\delta$ in \eq{wewant} cannot be
improved, which can be seen by considering a function of the form
$u(z,\theta)=v(\theta)/z$, with a nontrivial $v$ of vanishing
integral on $\partial M$, such that
$DDv+\frac{D^*Dv}{n-1}\hat{b}(0)\neq0$, where $D$ is the covariant
derivative operator of $\hat b(0)$, and such that $v$ is
$L^2(\partial M,\hat{b}(0))$-orthogonal to the kernel of
$P^*_{\hat{b}(0)}$ (see \eq{wewant2e} below).
\end{Remark}

 \proof
In some of the calculations of this proof the reader might find
it convenient to use the coordinate system of \eq{metrichgold}.
Without loss of generality we can assume that $\sigma\le 1$.
Let us define $d\nu_{\gdelta }=\delta^{n-1}d\mu_{\gdelta }$,
and note that the  measure $d\mu_{\gdelta }$ can be replaced by
$d\nu_{\gdelta }$ in \eq{orthoc}-\eq{orthocv}; e.g.,
\eq{orthoc} can be replaced by
\bel{orthoc1}
 \forall\ \mu=0,\ldots,k \qquad \int_A e^{-s/y}u u_{(\mu)}d\nu_{\gdelta }=0
 \;.
\ee
Suppose that \eq{wewant} with $d\mu_{\gdelta }$ there replaced by
$d\nu_{\gdelta }$ does not hold, then there exist sequences
$\delta_n \to 0$, $g^{(n)}$ and $u_n$ satisfying \eq{orthoc}
(respectively \eq{orthocv}) such that the right-hand-side  equals
one, while the reverse inequality to \eq{wewant} holds with $c_1$
replaced by $1/n$:
\beal{wewant2}
 &
 \displaystyle
  \int_{A} e^{-s/y}y^8|P^*_{\delta_n}
u_n|^2_{\gn }d\nu_{\gn}\leq \frac {\dnte }{n}
 \;,
 &
 \\
 &
 \displaystyle
\int_{A} e^{-s/y}(y^8|\nabla \nabla u_n|^2_{\gn }+y^4|\nabla
{u_n}|^2_{\gn }+u^2_{n})d\nu_{\gn }=1
 \;,
 &
\eeal{wewant3}
%
where we have set
$$
 \gn:= g^{(n)}_{\delta_n} \;\;\mbox{ and }\;\;P^*_{\delta_n}\equiv P^*_{\gn}.
$$
Let $y$ be the function on $A$ defined  in \eq{ydef}. Using
\eq{Pdelegg1} to express $\nabla \nabla u_n$ in terms of $P_{\gn
}^*u_n $ and $u_n$ one obtains (compare \eq{TSineq} below)
\bean
  \int_{A} e^{-s/y}y^8|\nabla\nabla
u_n|^2_{\gn }d\nu_{\gn }
 &\leq
 &
 \displaystyle C
  \int_{A} e^{-s/y}\Big(y^8|P^*_{\delta_n} |^2_{\gn }+u^2_{n}\Big)d\nu_{\gn }
 \\
 &
 \leq
 &
 \displaystyle C\left(\frac {\dnte }{n}+
  \int_{A} e^{-s/y} u^2_{n} \right)d\nu_{\gn }\;,
\eeal{wewant2a}
%
which together with \eq{wewant3} implies that there exists $c>0$
such that
\beal{wewant3a}
 &
 \displaystyle
 \int_{A} e^{-s/y}( y^4|\nabla {u_n}|^2_{\gn }+u^2_{n})d\nu_{\gn }\ge
 c
 \;.
 &
\eea
%
Now,
\beal{norm}
 &
 |\nabla u_n|_{\gn }^2 = z^2 \left(|\partial_zu_n|^2 +
 {\delta_n^2} |\partial_\theta u_n|_{{\hgn}  }^2
 \right)
 \;,
 &
\eea
where $|\cdot |_{{\hgn}  }$ denotes the norm of a tensor field on
$\partial M$ with respect to the metric
$$
{{\hgn}  }(z):= \hat {g}^{(n)}(\delta_n z)
 \;.
$$
Note that,  {decreasing the constant $\rho_0$ of Definition~\ref{D1}
if necessary}, all the ${{\hgn}  }$'s are uniformly equivalent to
$\hat b(0)$. {}From \eq{wewant3a} we obtain
\beal{wewant3b}
 &
 \displaystyle
 \int_{A} e^{-s/y}( y^4|\partial_z {u_n}|^2 +u^2_{n})d\nu_{\gn }\ge
c
 \;,
 &
\eea
for some $c>0$.

Clearly the trace of $P^*_{\gn } u$ satisfies an estimate of the
form \eq{wewant2} (compare Appendix~\ref{ASc})
\beal{wewant2c}
 &
 \displaystyle
  \int_{A} e^{-s/y}y^8|\Delta_{g_n}
u_n- nu_n+O(\delta_n^\sigma)u_n|^2 d\nu_{\gn }\leq C\frac {\dnte
}{n}
 \;.
 &
\eea
Let
$$
 \Esym:= \nabla\nabla u_n - \frac {\Delta_{\gn } \!\!\;u_n}n \,{\gn }=
  \nabla\nabla u_n - u_n \,\gn  + \mbox{ error }
\:,
$$
where the error term is bounded,  after integration, as in
\eq{wewant2c}. {}From  \eq{wewant2} and \eq{wewant2c} we
conclude that
\beal{wewant2d}
 &
 \displaystyle
  \int_{A} e^{-s/y}y^8|\Esym+O(\delta_n^\sigma)u_n |^2_{\gn } d\nu_{{\gn }}\leq C\frac {\dnte }{n}
 \;.
 &
\eea
Since $\Esym$ is trace-free we have
$\Esym_{zz}=-\dn^{-2}\hgn^{CD}\Esym_{CD}$, so that
\bel{norm2}
 |\Esym|_{\gn } ^2= z^4 \Big((1+\frac{\delta_n^4}{n-1})|\Esym_{zz}|^2 + 2 \delta_n^2|\Esym_{z A}|_{{\hgn} }^2
 + \delta_n^4 |\Esym_{AB}-{\hgn }^{CD}\Esym_{CD} {\hgn} {}_{AB}|_{\hgn }^2
 \Big)
 \;,
\ee
which together with the formulae in Appendix~\ref{ASc} (recall
we have assumed $\sigma<1$) leads to
\beal{wewant2e} &&
\\
\nonumber
 \displaystyle
  \int_{A} e^{-s/y}y^8\Big( |\partial_z^2u_n + z^{-1} \partial_z u_n - z^{-2}u_n+O(\delta_n^\sigma)u_n|^2
    \phantom{xxxxxxxx}
    &&
    \\
    \nonumber
  + \dn^2|\partial_z\partial_A u_n+ z^{-1}\partial_A u_n + O(\dn
  \partial_\theta
  u_n)+O(\delta_n^\sigma)u_n|_{{{\hgn} }}^2
  \phantom{xxxxx}
  \\
  \nonumber
 + \delta_n^4 |D_{A}D_{B}u_n-\hat g^{(n)}(\dn)^{CD}D_{C}D_{D}u_n ({\hgn }){}_{AB}
 +O(\delta_n^\sigma)u_n|_{{{\hgn} }}^2
 \Big)d\nu_{{{\gn }}}
   & \leq &
  C\frac {\dnte }{n}
 \;.
 \nonumber
\eea
Next, \eq{wewant2} together with the formula for $(P_\delg
^*u)_{zz}$ in Appendix \ref{ASc}  gives
\beal{wewant2f2} &&
\\
\nonumber
 \displaystyle
   \int_{A} e^{-s/y}y^8\Big|
     \left[(n-1)z+O(\delta_n^\sigma)\right]\partial_z
       u_n+\left[(n-1)+O(\delta_n^\sigma)\right]u_n
\phantom{xxxxxxxx}
    &&
    \\
    \nonumber
   -\delta_n^2\Delta_{\hat
g_{\delta_n}}u_n \Big|^2 d\nu_{{{\gn }}}\leq C\frac {\dnte }{n}
\;.
 \nonumber
\eea
Choose  $\delta_{n_0}\ne 0$  and let $H^1$ and $H^2$ be the Hilbert
spaces with   norms defined by the left-hand-sides of \eq{wewant3a}
and \eq{wewant3} with $n=n_0$, and   norms, covariant derivatives
and measures related to $b_{\delta_{n_0}}$:
\beal{normH1}
 &
 \displaystyle
\|u\|_{H^1}:= \int_{A} e^{-s/y}( y^4|\nabla {u
}|^2_{b_{\delta_{n_0}}}+u^2)d\nu_{b_{\delta_{n_0}}}
 \;,
 &
 \\
 &
 \displaystyle
\|u \|_{H^2}:= \int_{A} e^{-s/y}(y^8|\nabla \nabla u
|^2_{b_{\delta_{n_0}}}+y^4|\nabla {u
}|^2_{b_{\delta_{n_0}}}+u^2)d\nu_{b_{\delta_{n_0}}}
 \;.
 &
\eeal{normH2}

Now, \eq{wewant3} shows that $u_n$ and $y^2\partial_z u_n$ are
bounded in $L^2=L^2(A,e^{-s/y} d\nu_{b_{\delta_{n_0}}})$. Equation
\eq{wewant2c} proves that $y^4$ times the Laplacian of $u_n$ is
bounded in $L^2$. Further, \eq{wewant2e} establishes that
$y^4\partial^2_zu_n$ is bounded in $L^2$. Simple algebra gives then
that $y^4$ times the tangential Laplacian of $u_n$ is bounded in
$L^2$. Coming back to \eq{wewant2e} we obtain that all tangential
derivatives of $u_n$ are $L^2$-bounded, when multiplied by  relevant
powers of $y$. Standard interpolation gives an $L^2$-bound for $y^2$
times the first tangential derivatives of $u_n$. But \eq{wewant2e}
shows now that the functions $y^4\partial_z\partial_Au_n$ are
$L^2$-bounded. Finally, an interpolation will bound every (weighted)
first derivatives of $u_n$.

So, the sequence $u_n$ is bounded in $H^2$, therefore there exists a
subsequence, still denoted by $u_n$, which converges strongly in
$H^1$. But \eq{wewant2a} with $u_n$ replaced by $u_n-u_m$ shows that
$u_n$ is Cauchy in $H^2$, hence there exists $u\in H^2$ such that
$u_n$ converges to $u$ in $H^2$. {}From \eq{wewant2e} we infer that
\bea
 \nonumber
 \displaystyle   |\partial_z^2u  + z^{-1} \partial_z u  - z^{-2}u |^2
  +  |\partial_z\partial_A u + z^{-1}\partial_A u   |_{b_{\delta_{n_0}}}^2
  \phantom{xxxxx}
  \\
 +  |D_{A}D_{B}u -\hat b(0)^{CD}D_{C}D_{D}u  \,\hat b(0){}_{AB}|_{b_{\delta_{n_0}}}^2
   & = & 0
 \;,
 \eeal{ww3}
while  \eq{wewant2f2} implies
\bel{ww4} z\partial_zu+u=0.\ee
%
Solving \eq{ww3}-\eq {ww4}, we conclude that $u$ is a linear
combination of the $u_{(\mu)}$'s as given by \eq{KID3} for a
standard hyperbolic metric, while $u= \mbox{\rm const} /z$
otherwise.
But the integral in \eq{orthoc1} is continuous on $H^2$, which
implies that \eq{orthoc} is satisfied in the limit. Similarly,
$\delta^{n} $ times the integral \eq{orthocv} is continuous on
$H^2$. Recalling that the family $\{u_{(\mu)}\}$ is orthogonal with
respect to the scalar product defined by the integral in
\eq{orthoc}, we obtain $u=0$. This contradicts \eq{wewant3a}, and
proves the result. \qed

\medskip

Let $\psi=e^{-s/2y}$, $\phi= y^2$. We will use spaces
$\Hkd\equiv H^{k}_{\gdelta }$ of tensor fields on $A$
(compare~\cite{ChDelay}) for which the norms
 \be \label{defHn}
 \|u\|_{\Hkd }:=
\left(\int_A(\sum_{i=0}^k \phi^{2i}|\nabla^{(i)} u|^2_\gd)\psi^2
\delta^{(n-1)}d\mu_\gd\right)^{\frac{1}{2}}
 \ee
are finite, where $\nabla^{(i)}$ stands for the tensor
$\underbrace{\nabla ...\nabla }_{i \mbox{ \scriptsize times}}u$,
with $\nabla$ --- the Levi-Civita covariant derivative of $\gd$; we
assume throughout that the metric is at least $W^{1,\infty}_\loc$;
higher differentiability will be usually indicated whenever needed.
The factor $\delta^{(n-1)}$ in front of the measure $d\mu_{\gdelta
}$ has been included so that $\delta^{(n-1)} d\mu_{\gdelta }$ is
equivalent to the Lebesgue coordinate measure $dz d\theta$,
uniformly in $\delta$.

Note that $H^0_\gdelta$ involves weights, but  $L^2$ does not.

An equivalent norm, and therefore the same space, is obtained
if $\gdelta $ in \eq{defHn} is replaced by $\bdelta $.

We will need the following:

\begin{Lemma}
Let  $c_0,\sigma>0$ and  $s\ge 0$. There exist constants
$C=C(n,\ell,s,c_0,\sigma)>0$ and
$\delta_0=\delta_0(n,\ell,s,c_0,\sigma)>0$ such that for all
$(c_0,\ell+2,\sigma)$-asymptotically hyperbolic metrics $g$ and
for all $0<\delta\leq\delta_0$
 \label{Lhdes}
$$
 \|u \|_{H^{\ell+2}_{\gdelta }} \le C \Big( \|\phi^2 P_{\gdelta } ^*u \|_{{H^{\ell}_{\gdelta }}} +  \|u
 \|_{{H^{0}_{\gdelta }}}\Big)
 \;.
$$
\end{Lemma}

\proof For $\ell=0$ the result has  been established in the
course of the  proof of Proposition~\ref{Pcontr}, see the first
line of \eq{wewant2a}.  For $\ell=1$ we start the calculation
that follows with $k=2$ and we stop at the second line,
invoking weighted interpolation and the result for $\ell=0$ to
conclude. Otherwise, suppose that the result is true for
$k-1\le \ell_0$ with $\ell_0\ge 1$.
Using~\cite[Equation~(A.4)]{ChDelay}

(one can check that the constants in equations (A.2) and (A.3)
there, thus also in (A.4),  do  not depend on $\delta$) to
control the first term when passing from the second to the
third line below, we find for $2\le k-1+2\le \ell_0+2$
\beaa
 \|\phi^{k+1}
 \nabla^{(k-1)} (\nabla^{(2)} u - \Delta u
 \gdelta)\|_{H^0_{\gdelta }}
 & = &
  \| \phi^{k+1} \nabla^{(k-1)} (P^*u
-(n-1) u+O( \delta ^\sigma)u)\|_{H^0_{\gdelta }}
 \\ & \le &
  \| \phi^{k+1} \nabla^{(k-1)}  P^*u \|_{H^0_{\gdelta }}\\
  &&\;\;+
  C_1\|\underbrace{\phi^2}_{\le C} \phi^{k-1} \nabla^{(k-1)}
[(1+O(\delta^\sigma))u]\|_{H^0_{\gdelta }}
 \\
 & \le  &
 C\Big(
  \| \phi^2 P^*u\|_{H^{k-1}_{\gdelta }}
+C_2\underbrace{ \|u\|_{H^{k-1}_{\gdelta }}}_{\le C  (\|  \phi^2
P^*u\|_{H^{k-3}_{\gdelta }}+ \|u\|_{H^0_{\gdelta }})}
 \Big)
 \\
 & \le  &
  C\left(1+CC_2\right)\| \phi^2  P^*u\|_{H^{k-1}_{\gdelta }}
+C^2C_2 \|u\|_{H^0_{\gdelta }}
 \;.
\eeaa
This is the desired inequality, to see this set
$$
T:= \nabla^{(k+1)} u
 \;,
  \qquad
  S:= \nabla^{(k-1)} (\nabla^2 u - \Delta u \gd)
 \;,
$$
or, in index notation,
$$
T_{i_{1}\ldots i_{k-1}jk}:= \nabla_{i_1}\cdots
\nabla_{i_{k-1}}\nabla_j\nabla_k u $$
$$
  S_{i_{1}\ldots i_{k-1}jk}:= \nabla_{i_1}\cdots
\nabla_{i_{k-1}}\nabla_j\nabla_k u - \nabla_{i_1}\cdots
\nabla_{i_{k-1}}\nabla^\ell\nabla_\ell u \;(\gd)_{jk}
 \;,
$$
 straightforward algebra shows that
\bel{TSineq}
 |T|^2_\gd \le  |S|^2_\gd
 \;,
\ee
and the Lemma follows.
\qed

\bigbreak

As in~\cite{ChDelay} we set
\bel{Ldelta}
 L_{\gdelta }:= \psi^{-2} P_{\gdelta } \phi^4 \psi^2P_{\gdelta }^*
 \;,
\ee and \bel{Kdelta} \KDE=\ker P^*_{\bdelta }\;.\ee
The proof of~\cite[Theorem~3.6]{ChDelay} shows that
$$
 L_{\gdelta }^{-1}:
 H^k_{\gdelta }\cap K_\delta^{\perp_{H^0_{\gdelta }}} \to H^{k+4}_{\gdelta }
$$
exists for $\delta$ small enough. However,   uniform boundedness in
$\delta$ of $L_{\gdelta }^{-1}$ does not hold in our case, instead
we have:

\begin{Corollary}\label{invLb}
Let $k\in\N$,  $c_0,\sigma>0$ and  $s\ge 0$. There exist
constants
$C=C(n,k,s,c_0,\sigma)>0$ and
$\delta_0=\delta_0(n,k,s,c_0,\sigma)>0$ such that for all
$(c_0,k+4,\sigma)$-asymptotically hyperbolic metrics $g$, for
all $0<\delta\leq\delta_0$, and for all $u$ satisfying
\eq{orthoc} or \eq{orthocv}
\label{Cest}
$$
 \|L_{\gdelta }^{-1} u \|_{H^{k+4}_{\gdelta }} \le C\Big(
 \|u\|_{H^{k}_{\gdelta }}+\delta^{-4}
 \|u\|_{H^0_{\gdelta }}\Big)
 \;.
$$
\end{Corollary}

\proof By Proposition~\ref{Pcontr} we have  (recall that $H^0$
is weighted but  $L^2$ is not)
\beaa
c\delta^4 \|u\|_{H^2_{\gdelta }}^2 &\le&  \|\phi^2P_{\gdelta }^*u \|_{{H^0_{\gdelta }}}^2 \\
 & = &
\langle \psi^2 \phi ^2 P_{\gdelta } ^*u
 , {  \phi^2  P_{\gdelta } ^*}  u \rangle_{L^2_{\gdelta }}
 =
 \langle \psi^2u
 ,\underbrace{\psi^{-2}P_{\gdelta } \phi^4 \psi^2 P_{\gdelta } ^*}_{L_{\gdelta }} u \rangle_{L^2_{\gdelta }}
 \\
 & = &
 \langle \psi u ,\psi L_{\gdelta } u\rangle _{L^2_{\gdelta }} \le \|\psi L_{\gdelta  u} \|_{{L^2_{\gdelta }}}
 \|\psi u\|_{{L^2_{\gdelta }}}= \|L_{\gdelta } u \|_{{H^0_{\gdelta }}}
 \|u\|_{{H^0_{\gdelta }}}
 \;.
 \eeaa
Replacing  $u$ by $ L_{\gdelta } ^{-1}u$ we conclude that
\bel{presque} c\delta^4 \|L_{\gdelta }^{-1}u\|_{H^2_{\gdelta }} \le
\| u \|_{H^0_{\gdelta }}
 \;.
 \ee
In order to finish the proof we will use the following elliptic
estimate, which is standard except for the uniformity in $\delta$;
the proof can be found in Appendix~\ref{A1}:

\begin{Lemma}
 \label{Lhdes2}
Under the conditions of Corollary~\ref{invLb}, there exists a
constant $C$, independent of $g$  and $\delta$, such that for
$\delta$ small
\bel{hw1}
 \|u \|_{H^{k+4}_{\gdelta }} \le C \Big( \|L_{\gdelta } u \|_{H^{k}_{\gdelta }} +  \|u
 \|_{H^{0}_{\gdelta }}\Big)
 \;.
\ee
\qed
\end{Lemma}

Returning to the proof of Corollary~\ref{Cest}, we replace $u$
by   $ L_{g_\delta} ^{-1}u$  in \eq{hw1} to obtain
\bel{hw1bis}
 \|L_{\gdelta } ^{-1}u\|_{H^{k+4}_{\gdelta }} \le C \Big( \|u \|_{H^{k}_{\gdelta }} +  \|L_{\gdelta } ^{-1}u
 \|_{H^{0}_{\gdelta }}\Big)
 \;,
\ee
and the Corollary follows from \eq{presque}.
 \qed

Summarizing, we have proved:

 \begin{theorem}\label{invLg}
Let $k\in\N$, $\sigma>0$, $c_0>0$ and  $s\ge 0$. There exist
constants $C=C(n,s,\sigma,c_0)>0$ and $\delta_0=\delta_0(n,s
,\sigma,c_0)>0$ such that for all $(c_0,k+4,\sigma)$-asymptotically
hyperbolic metrics $g$, for all $0<\delta\leq\delta_0$ and  for any
$u\in H^{k+4}_{\gdelta }\cap \KDE^{\bot_{\gdelta }}$,
$$
C\delta^4||u||_{H^{k+4}_{\gdelta }}\leq ||L_{\gdelta
 }u||_{H^{k}_{\gdelta }}\;.
$$
In particular the operator $\Pi_{\KDE^{\bot_{\gdelta }}}L_{\gdelta
}$, where $\Pi_{\KDE^{\bot_{\gdelta }}}$ denotes orthogonal
projection
  on ${\KDE^{\bot_{\gdelta }}}$ in $H^0_{\gdelta }$, is an
isomorphism from $H^{k+4}_{\gdelta }\cap \KDE^{\bot_{\gdelta }}$ to
$H^{k}_{\gdelta }\cap \KDE^{\bot_{\gdelta }}$ such that the norm of
its inverse is bounded by $C^{-1}\delta^{-4}$.
\end{theorem}

 \qed

At this point, we have established Step 2) of the Introduction,
as well as some elements of   Step 3). We continue with further
details of Step 3).

\section{The gluing construction on a moving annulus}
 \label{Sgcma}

In this section we prove Theorem~\ref{Tm2intro}. We  set
$k=\ell-4$. We consider conformally compact asymptotically
hyperbolic metrics $g$ which asymptote, with $k+2$ derivatives,
to a fixed AH metric $b$. We fix a small $\delta_0>0$ and
define the space $W^{k+4,\infty}_b(M_{4\delta_0})$ of symmetric
two tensors with $k+4$ $b$-covariant derivatives bounded on
$M_{4\delta_0}$, relatively to the norm of $b$.
Following~\cite{ChDelay}, we assume that $g-b$ is close to zero
in $W^{k+4,\infty}_b(M_{4\delta_0})$.

Similarly to \eq{bdelmet}, we denote by $\gdelta $ the metric on
$A_{1,4}$ obtained by restricting $g$ to $A_{\delta,4\delta}$, and
rescaling the $\rho$ coordinate to $A_{1,4}$. Unless explicitly
specified otherwise, covariant derivatives on $A_{1,4}$ are related
to $\gdelta $.

As in~\cite{ChDelay}, consider the map
$$
\begin{array}{llll}
\Fd: & \psi^2\phi^2 H^{k+2}_\delta& \longrightarrow & H^k_\delta\cap \KDE ^\bot\\
 &h&\longmapsto&\Pi
 \{\psi^{-2}[R(\gdelta +h)-R(\gdelta)]\},
\end{array}
$$
where $\KDE = \Ker \Pbd^* $, where  $\Pbd^* $ is as in
\eq{Pdeleq}, 
and $\Pi$ is the $H^0_\delta$ projection onto $\KDE ^\bot$, the
$H^0_\delta$-orthogonal of $\KDE $; all the spaces here are spaces
of tensors on $A_{1,4}$.

One should keep in mind that we are interested in $h$'s of the
form $h=\psi^{2}\phi^4P^*u$, $u\in H^{k+4}_\delta \cap
 \KDE ^\bot$, with $u$ small in the last space.

Near $h=0$ the map $\Fd $ is a smooth map between Hilbert
spaces. We consider now~\cite[Proposition G.1]{ChDelay} with
$x=\gdelta $ so that $f_x$ there equals $\Fd $ here. One checks
that $f$ satisfies conditions (2) and (3) of~\cite[Proposition
G.1]{ChDelay} with the set $A$ there being
\bel{Aset}
 A=\{ \gdelta \;,\ 0<\delta \leq \delta_0\;,\ (g-b) \mbox{ sufficiently small
in } W^{k+4,\infty}_b(M_{4\delta_0})\}
 \;. \ee
Furthermore,
$$
V_x=\psi^2\phi^2H^{k+2}_\delta,\;\; W_x=H^{k}_\delta\cap \KDE ^\bot
 \;.
$$
Now, $\Fd $ satisfies a modified version of  condition (1) there:
here, by Theorem~\ref{invLg}, we have that $Df_x(0)$ has a right
inverse $\psi^2 \phi^4 P^*_{\gdelta } L_{\gdelta }^{-1}$ bounded by
$C_1\delta^{-4}$, where $C_1$ does not depend upon $x\in A$. For the
sake of notational legibility we present the argument \emph{without}
using the smoothing operators of~\cite{ChDelayHilbert}; the latter
provide what is needed to obtain the  differentiability claimed.
Note that we haven't assumed any uniformity in $\delta$ on the
modulus of the H\"older continuity of $g$, as the solution will
exist, and will have H\"older regularity, without any such
assumptions. Any further hypotheses about uniformity of that modulus
would be reflected in associated uniformity for the metrics obtained
by the gluing procedure, but such uniformity is irrelevant for our
purposes.

A repetition of the proof of Proposition~G.1 of~\cite{ChDelay} with
$C_1$ there replaced with $C_1\delta^{-4}$ yields:

\begin{theorem}\label{solmodker}
There exist constants $\epsilon>0$ and $C>0$ such that for all
$\delta$ sufficiently small and for all functions $f\in H^k_\delta $
with
$$
||f||_{H^k_\delta}\leq \epsilon\delta^4,
$$
there exists a unique $h=\psi^{2}\phi^4P^*_{\gdelta }u$, with
$||\psi^{-2}\phi^{-2}h||_{H^{k+2}_\delta }$ close to zero,
satisfying $\Fd (h)=\Pi f$ and
$$
||\psi^{-2}\phi^{-2}h||_{H^{k+2}_\delta}\leq
C'||u||_{H^{k+4}_\delta}\leq C \delta^{-4} ||f||_{H^k_\delta}\leq
C\epsilon.
$$

\end{theorem}
We will use  Theorem~\ref{solmodker} to glue an AH metric $g$, with
timelike energy-momentum vector, with a Kottler one $b_{p_{(\mu)}}$,
on an annulus $A_{\delta,4\delta}$. Let $\chi$ be a cutoff function
equal to zero on $A_{1,2}$ and to one on $A_{3,4}$. We define a
first glued metric on $A_{1,4}$ as
\bel{glue0}
 g_{\delta,p_{(\mu)}}:=\chi \gdelta +(1-\chi)b_{\delta,p_{(\mu)}}
 \;.
\ee
It is clear that the metric $g_{\delta,p_{(\mu)}}$ belongs to the
set $A$ of \eq{Aset}. Set
\bel{fdef}
 f:=
 \psi^{-2}[R(\bdelta)-R(g_{\delta,p_{(\mu)}})]=
 \psi^{-2}[R(b_{\delta,p_{(\mu)}})-R(g_{\delta,p_{(\mu)}})]
 \;.
\ee
Let $\pzm$ be the momentum vector of $g$. We will assume that $g$
has
  the following asymptotic behaviour
\bel{condg} |g-\bpzm|_b+|\zD( g-\bpzm)|_b+\ldots
+|\zD^{(k+2)}(g-\bpzm)|_b=O(\rho^\alpha)
 \;,
\ee
for some $\alpha>0$, to be restricted shortly; here $\zD$ is the
covariant derivative of $b$. Recall that
\bel{condb} |b_{p_{(\mu)}}-b|_b+|\zD b_{p_{(\mu)}}|_b+\ldots
+|\zD^{(k+2)}b_{p_{(\mu)}}|_b=O(\rho^n). \ee
Under   \eq{condg}  we have
\bel{condgdp} |g_{\delta,p_{(\mu)}}-\bdpzm|_{\bdelta }+|\nabla(
g_{\delta,p_{(\mu)}}-\bdpzm)|_{\bdelta }+\ldots
+|\nabla^{(k+2)}(g_{\delta,p_{(\mu)}}-\bdpzm)|_{\bdelta
}=O(\delta^\alpha)
 \;,
\ee
where the norm and covariant derivatives are defined  by $\bdelta $.
This implies
\beal{condgdp2} |g_{\delta,p_{(\mu)}}-\bdpm|_{\bdelta }
 &\le&
 |g_{\delta,p_{(\mu)}}-\bdpzm|_{\bdelta }
 + |\bdpm-\bdpzm|_{\bdelta }
 \\
 \nonumber
 &= &O(\delta^\alpha)+
O(|\pgm-\pzm|\delta^n)
 \\
 \nonumber
 &=&
 O(\delta^\alpha)+
 O(\delta^{\beta+n})
 \;,
\eea
provided that $\pgm$ is assumed to satisfy
\bel{prest} |\pgm-\pzm|  =O(\delta^{\beta})
 \;,
\ee
for some $\beta>0$. An inequality similar to \eq{condgdp2}
holds for derivatives of order up to $k+2$. It follows that the
function $f$ defined in \eq{fdef} satisfies
$$
||f||_{H^k_\delta}=O(\delta^\alpha) +
 O(\delta^{\beta+n})\;.
$$
By Theorem~\ref{solmodker}  if
\bel{firstcond1}
 \alpha>4\;, \quad \beta+n>4
 \;,
\ee
then for all $\delta$ small enough there exists a solution
$h_{\delta,p_{(\mu)}}$ to the equation
$$
 f_{g_{\delta,p_{(\mu)}}}(h_{\delta,p_{(\mu)}})=\Pi f
 \;,
$$
with
\bel{lastest}
 ||\psi^{-2}\phi^{-2}h_{\delta,p_{(\mu)}}||_{H^{k+2}_{\delta}}=O(\delta^{\alpha-4})+
 O(\delta^{\beta+n-4})
 \;.
\ee
 Summarizing, for all $p_{(\mu)}$,  we have constructed a solution
$h_{\delta,p_{(\mu)}}$, modulo kernel, to the equation
$$
 R(g_{\delta,p_{(\mu)}}+h_{\delta,p_{(\mu)}})-R(b)=0
 \;,
$$
satisfying \eq{lastest}.  This finishes Step 3) of the
Introduction.

We now proceed to Step 4). Set
$$
 \widetilde{g}_{\delta,p_{(\mu)}}=g_{\delta,p_{(\mu)}}+h_{\delta,p_{(\mu)}}
 \;.
$$
We consider now the projection onto the kernel as follows. For all
$\delta$ small, and for all $p_{(\mu)}$ satisfying \eq{prest}, we
define
\bel{Ideltadef}
 I_\delta(p_{(\mu)})=\frac{1}{\delta^n}\pi[\psi^{-2}(R(\widetilde{g}_{\delta,p_{(\mu)}})-R(\bdelta))]
 \;,
\ee
where $\pi$ is the $H^0_\delta$ orthogonal projection onto
$\KDE $.  We want to show that we can choose $p_{(\mu)}$ such
that $I_\delta(p_{(\mu)})=0$. We need the following identity,
from~\cite{ChHerzlich}:
\begin{eqnarray}
  \label{eq:3.2} &
  \sqrt{\det g} \;\sqrtV (R_g-R_b)  =  \partial_i \left(\ourU^i(\sqrtV)\right) + \sqrt{\det g}
\;(  \rho + Q)\;, &
\end{eqnarray}
where
\begin{eqnarray} \label{eq:3.3} & {}\ourU^i (\sqrtV):=  2\sqrt{\det
g}\;\left(\sqrtV g^{i[k} g^{j]l} \zD_j g_{kl}
+D^{[i}\sqrtV  
g^{j]k} e_{jk}\right)
\;,&
\\
\label{eq:3.4} & \rho:= (-\sqrtV \Ric(b)_{ij} +\zD_i \zD_j \sqrtV
-\Delta_b \sqrtV  b_{ij}) g^{ik} g^{j\ell}
  e_{k\ell}
\;,&
\\
\label{eq:3.5} & Q:= \sqrtV (g^{ij} - b^{ij} + g^{ik}g^{j\ell}
e_{k\ell})\Ric(b)_{ij} +Q'\;.
\end{eqnarray}
Brackets over a symbol denote anti-symmetrisation, with an
appropriate numerical factor ($1/2$ in the case of two
indices), and $\zD$ denotes the covariant derivative operator
of the metric $b$; note that $\rho$ here should not be confused
with the defining function of the boundary. Here $ Q'$ denotes
an expression which is bilinear in
$$e\equiv e_{ij}dx^i dx^j
:= (g_{ij}-b_{ij})dx^i dx^j
 \;,
$$
and in $\zD_k e_{ij}$,  linear in $\sqrtV $, $d\sqrtV $ and
Hess$\sqrtV $, with coefficients which are constants in  any ON
frame for $b$. The key is that $\rho$ vanishes when $\sqrtV$ is
in the kernel of $P^*_b$, and then $Q$ is at least quadratic in
$e$ near $e=0$. Indeed, the first term at the right-hand-side
of \eq{eq:3.5} does  not contain any terms linear in $e_{ij}$
when Taylor expanded at $g_{ij}=b_{ij}$.


The integral of $\ourU$ at the boundary at infinity provides
the momentum vector, and we need to know how fast the limit is
approached. The simplest case arises when $g$ is a Kottler
metric $b_m$ with mass parameter $m$, so that  (see
\eq{bmetinr} and \eq{bmetinr2})
\bel{gbeq} b_m= \frac {dr^2} {W^2} + r^2 \zh\;, \quad b= \frac
{dr^2} {\zW^2} + r^2 \zh
 \;,
\ee
where $\zh$ is the unit round metric on the sphere $\Sp^{n-1}$.
If $\{r\}\times \Sp^{n-1}$ is positively oriented,
a calculation gives
\bel{massform}
 \int_{\{r\}\times \Sp^{n-1}} \ourU^i dS_i = 2\omega_{n-1} (n-1)
 \zW W^{-1}m = 2\omega_{n-1} (n-1)
  m + O(r^{-n})
  \;,
\ee
where $\omega_{n-1}$ is the volume of $\Sp^{n-1}$. An identical
formula, with $\omega_{n-1}$ replaced by the $\hat b$--volume of
$N$, holds for the non-spherical Kottler metrics \eq{bKottler}.
Next, assume that $|g-b|_b=O(r^{-\alpha})$, where $r$ is a
coordinate for $b$ as in \eq{gbeq}, and $\zh$ is an
($r$--independent) metric on the compact conformal boundary $N$,
with the same decay rate for first derivatives, and with
$R(g)=R(b)$. Integrating \eq{eq:3.2} over $[r,\infty)\times \bmanN$
one finds, for $\alpha> n/2$,
\bel{decaygen}
 \int_{\{r\}\times \bmanN} \ourU^i(\Vm) dS_i = \pgm +
 O(r^{n-2\alpha}) \;,
\ee
which coincides of course with \eq{massform} if $\alpha=n$; we note
that $\alpha=n$ is the appropriate rate for Kottler metrics, whether
boosted or not. 

To calculate \eq{Ideltadef} explicitly, let $\Vmd$ be a basis of
$\KDE$, the vanishing of \eq{Ideltadef} is then equivalent to the
vanishing of the collection of integrals
$J_\delta(p)=(J_{\delta,(\nu)})$, where $p=(\pgm)$ and
\begin{eqnarray*} J_{\delta,{(\nu)}}&:=&
\int_{A_{1,4}}\psi^2\psi^{-2}(R(\widetilde{g}_{\delta,p_{(\mu)}})-R(\bdelta
))\Vnd \dmud \;.
\end{eqnarray*}
In order to use \eq{eq:3.2} we need to change the measure
$\dmud$ to $\dmutgd$, the following estimates are useful for
that:
$$
R(\widetilde{g}_{\delta,p_{(\mu)}})-R(\bdelta) = O(\delta^\alpha)+
O(\delta^{n+\beta})\;,$$
$$
\dmud=\left(1+O(\delta^{\alpha-4})+
O(\delta^{n+\beta-4})+O(\delta^n)\right)\dmutgd\;.$$
Keeping in mind that $\Vnd$ behaves as $\delta^{-1}$, and that the
volume form grows as $\delta^{1-n}$, this leads to
\begin{eqnarray*}
J_{\delta,{(\nu)}}
 &=& \int_{A_{1,4}}\psi^2\psi^{-2}(R(\widetilde{g}_{\delta,p_{(\mu)}})-R(\bdelta))\Vnd \dmutgd
\\
 &&
 +O(\delta^{2\alpha-4-n })
 +O(\delta^{2(\beta+n)-4-n})
+O(\delta^{\alpha})+O(\delta^{n+\beta})
\\
&=&p^0_{(\nu)}-p_{(\nu)}+O(\delta^{2(\alpha-4)-n })+O(\delta^{2(\beta+n-4)-n })+O(\delta^{n})+O(\delta^{2\alpha-n})\\
&=&\delta^{\beta}\Big(
\frac{p^0_{(\nu)}-p_{(\nu)}}{\delta^{\beta}}+O(\delta^{2\alpha-8-n-\beta})+O(\delta^{\beta
+n-8})+O(\delta^{n-\beta})+O(\delta^{2\alpha-\beta-n})\Big)
 \;.
\end{eqnarray*}
Here    the terms $O(\delta^{2(\alpha-4)-n})$ and
$O(\delta^{2(\beta+n-4)-n})$ in the third line arise from the
terms quadratic in \eq{eq:3.2} and from the first two terms in
the second line, while the terms $O(\delta^{n})$ and
$O(\delta^{2\alpha-n})$ arise from the difference between the
boundary term and its limit (namely $p^0_{(\mu)}-p_{(\mu)}$)
when $\delta$ goes to zero, compare \eq{massform} and
\eq{decaygen}, and also contain the last two terms in the
second line. To close the argument all error terms should go to
zero as $\delta$ tends to zero, thus
\bel{ineq}
 2\alpha-8-n-\beta>0\;,\quad {\beta
+n-8}>0\;,\quad {n -\beta}>0
 \;.
\ee
(Note that \eq{firstcond1} does not impose any further
restrictions.) This is equivalent to
\bel{ineq2} n\ge 5\;,\quad
 \alpha>\frac{8+n+\beta}2\;, \quad max(8-n,0)<{\beta}<n
 \;.
\ee
So $\beta$ can be chosen consistently with those bounds provided
that \eq{ineq2i} holds.

If the  kernel of $P^*_b$ is one-dimensional, with $\pzm=m_0$,
then for $\delta$ small, using  the intermediate value theorem,
there exists $\pzgm=m$ in an interval $[-m_0,2m_0]$  such that
$J_\delta(m)=0=I_\delta(m)$, proving existence of a solution.

Otherwise, under \eq{ineq2}, we can use   a Brouwer fixed point
theorem as in Lemma 3.18 of~\cite{ChDelay} with:
\begin{itemize}
\item  $U$: a bounded open ball of centre $0$ in
    $\R^{n+1}$;
\item $G=Id$: $\qgm\mapsto \qgm$,
\item $V=U$,
\item $\lambda = 1/\delta $ and $G_\lambda=G_{1/\delta}=\delta^{-\beta}J_\delta(\pzm+\delta^\beta \qgm)$,
\item $y=0$.
\end{itemize}
This shows that for small $\delta$, we can choose $p_{(\mu)}$
so that $I_\delta(p_{(\mu)})=0$, which again proves existence
of  a solution.

Regularity follows from~\cite[Theorem~4.9]{ChDelayHilbert}.
This completes the proof of Theorem~\ref{Tm2intro}.
 \qed

\section{The gluing construction on a fixed annulus  }
 \label{A7}

The question addressed in Theorem~\ref{TM1gN}   is a special case of
the following: In dimension $n\ge 3$, consider an $n$-dimensional
submanifold $M\subset \Mext$, where $\Mext$ has been defined in
\eq{Mext}, with compact, connected, nonempty boundary $\partial
\hyp$ which separates $\Mext$ into two components, one of which is
bounded. We further suppose that  $M$ is included in the  bounded
component, and that $\hyp$ is equipped with a metric $g\in C^{\ell
,\lambda }$, $\ell
>\lfloor \frac n 2 \rfloor +4$, $\lambda \in (0,1)$, of constant
negative scalar curvature.

Let $M_1\subset \hyp$ be a one-sided collar neighbourhood of
$\partial \hyp$ contained in $M$, we will refer to $M_1$ as the
\emph{interior collar}.

Let $M_2\subset \Mext$ be a one-sided collar neighbourhood of
$\partial M$ which lies in the unbounded component of $\Mext$,
we refer to $M_2$ as the \emph{exterior collar}.

The  \emph{extension problem} is to find a constant scalar curvature
metric on $M\cup M_2$ which coincides with  $g$  on $M$, and which
coincides with a Kottler metric  near $\partial M_2\setminus
\partial M$. In
this case we set $M'=M$ and $M''=M\cup M_2$. In this problem one
would presumably want $M_2$ to be small: a solution with a small
$M_2$ provides a solution for any bigger one.

The  \emph{deformation problem} is to find a constant scalar
curvature metric on $M $ which coincides with  $g$  on $M\setminus
M_1$, and which coincides with a Kottler metric  near $\partial M$.
In this case we set $M'=M\setminus M_1$ and $M''=M $.  Similarly to
the previous problem, one would presumably  want $M_1$ to be small.

Let $\pzm$ denote the energy-momentum vector of $\partial  M$,

defined as:
\bel{pint}
 \pzm = \int_{\partial  M} \ourU^i(V_{(\mu)}) dS_i \;,
\ee
with $\ourU$ as in \eq{eq:3.3}, and $V_{(\mu)}$ defined in
Section~\ref{sec:def}. If $(N,\hat b)$ is the round sphere, then
$\pzm$ is a vector in $\R^{n+1}$. Otherwise $\pzm$ is simply a
number, say $m_0$.

Denote by
$$
 \psi_{\partial M}: b_{\pgm}\mapsto \int_{\partial  M} \ourU^i(V_{(\mu)}) dS_i \;,
$$
the map which to a Kottler metric $b_\pgm$ associates the
energy-moment vector of  $\partial  M$, where $\ourU^i$ is
calculated using \eq{eq:3.3} with $g$ there replaced by $b_\pgm$.

\begin{Remark}
\label{Remarpsidm} As an illustration, assume that $\partial M =
\{r\}\times N$ for some $r$. Suppose  that $(N,\hat b)$ is
\emph{not} a round sphere, then $ \psi_{\partial M}^{-1}$ is a
smooth diffeomorphism between an interval of masses around $m_0$ and
its image; this follows immediately from the (non-spherical
equivalent of the) first equality in \eq{massform}. The result
remains true in the spherical case when one restricts
$\psi_{\partial M}$ to the standard, \emph{unboosted} Kottler
metrics $b_m$ as given by \eq{bmetinr}.
\end{Remark}

Similarly, consider $\partial M = \{r\}\times N$ for some $r$,
with $(N,\hat b)$ --- a round sphere, and assume moreover that
$\pzm$ lies in the image of $\psi_{\partial M}$. It is then
easily seen from \eq{decaygen} that $ \psi_{\partial M}^{-1}$
provides a smooth diffeomorphism between a neighbourhood of
$\pzm$ and its image \emph{provided that $r$ is large enough}.

We have the following:

\begin{Theorem}\label{TM1}
Let $n\geq 3$, $\N  \ni \ell>\lfloor \frac{n}{2}\rfloor + 4$,
$\lambda\in (0,1)$. Assume that the map $ \psi_{\partial
M}^{-1}$ is a homeomorphism of a neighbourhood of  $\pzm$ and
its image. There exists $\epsilon>0$ such that if
$$
 \|g-b_{\psi^{-1}(\pzm)}\|_{C^{\ell ,\lambda }(M)}<\epsilon
  \;,
$$
then there exists   a $C^{\ell ,\lambda }$ metric of constant
negative scalar curvature which coincides with $g$ on $ M'$, and
which is a Kottler metric on the unbounded component of $\Mext
\setminus \partial M$ away from $M''$. If $g$ is smooth, then so is
the solution of the deformation problem.
\end{Theorem}

\proof
We proceed as in~\cite[Section~8.6]{ChDelay} but we use the
refined versions of Theorem~5.6 and Proposition~5.7
of~\cite{ChDelay} used there, as given by Theorems~3.1 and  4.9
of~\cite{ChDelayHilbert} (compare Section~6.3
of~\cite{ChDelayHilbert}). The gluing is done on the collar
neighbourhood $  [0,1] \times \partial \hyp $, with $g_0$ there
being $b_{\pgm}$, $K=\delta J=0$, $g$ there equal to $\chi
g+(1-\chi)b_{p_{(\mu)}}$ where $\chi$ is a cutoff function
which vanishes near $\partial \hyp $, which we identify with
$\{1\}\times \partial \hyp$; finally, $p_{(\mu)}$ is close to
$\psi^{-1}_{\partial M}(\pzm)$. One thus obtains a solution
modulo kernel (note that for the estimates on pp. 53-54
of~\cite{ChDelay}, we have to replace $\rho$ there with
$R+n(n-1)$). For the kernel projection
(see~\cite[Equation~(8.24), p.~55]{ChDelay}) we proceed as
in~\cite{ChDelay}, p.~55, where $Q$ there is replaced by $p$
here. By~\cite[Lemma~3.3]{ChPollack} the kernel at
$b_{\psi^{-1}(\pzm)} $ is one-dimensional except in the
spherical case with $b_{\psi^{-1}(\pzm)}=b$, so except for this
last case this is a straightforward continuity argument by
varying masses in an interval around $m_0$. The H\"older
regularity of the final metric follows from Theorem~4.9
of~\cite{ChDelayHilbert}.
\qed

\bigskip

{\noindent \sc Proof of Theorem~\ref{TM1gN}:} The result
follows immediately from Theorem~\ref{TM1} and
Remark~\ref{Remarpsidm} except in the spherical case with
$b_{\psi^{-1}(\pzm)}=b$. In this last situation, the
supplementary hypothesis of parity insures that all
constructions can be made within the class of parity symmetric
metrics. The kernel within this class is one-dimensional, and
the solution can be adjusted by changing $b_m$ in the exterior
region within the family of unboosted Kottler metrics; compare
the proof of Theorem~2.1 in~\cite{ChDelay2}.
\qed

\section{$b$-conformal deformations near infinity}
 \label{sbdb}
Let $\bM $ be a compact manifold with boundary, set $M=\bM \setminus
\partial M$, and let $\rho$ be a defining function for $\partial M$.
Let $\bar b$ be a  $C^{k+2,\alpha}$ metric on $\bM $. Let $  h$
be covariant symmetric two tensor field such that $  g=b+ h$ is
positive definite, and for  functions $v>-1$ set $u=1+v$. For
$\delta>0$ and $h\in C^{k+2,\alpha}_\delta$ we consider the
function $F_{  g}$ defined on a neighbourhood of zero in
$C^{k+2,\alpha}_\delta$ to $C^{k,\alpha}_\delta$ as
$$
 F_{  g}(v)=-4\frac{n-1}{n-2}\nabla^k\nabla_ku+R(g)u+n(n-1)u^{\frac{n+2}{n-2}},
$$
where covariant derivatives are related to $g=b+h $, with the spaces
$C^{k,\alpha}_\delta$ of  tensors fields or functions as
in~\cite{Lee:fredholm}.  Note that $F_g( v)=0$ if and only if  the
scalar curvature of $u^{\frac{4}{n-2}}g$ equals $-n(n-1)$. The map
$F_{  g}$ is smooth near zero and, if $R(g)=-n(n-1)$, the derivative
at $v=0$ given by
$$
F'_{  g}(0)w=4\frac{n-1}{n-2}(-\nabla^k\nabla_k +n)w
 \;.
$$
%
The map {$F'_{  g}(0)$} is an isomorphism from
$C^{k+2,\alpha}_\delta$ to $C^{k,\alpha}_\delta$ when
$\delta\in(-1,n)$~\cite[Theorem~7.2.1]{AndChDiss}, so in
particular when $\delta\in(0,n)$. The implicit function theorem
then shows:

\begin{proposition}\label{scalconf}
Let $k\in \N$, $\delta\in(0,n)$, $\alpha \in (0,1)$, and for
$\zh \in C^{k+2,\alpha}_\delta$ let $ \mathring g=b+\zh $ be a
metric on $ M $ as described above with constant scalar
curvature $-n(n-1)$. There exists $\epsilon>0$ and a constant
$C$ such that for any $h\in C^{k+2,\alpha}_\delta$ with norm
less than $\epsilon$ there exists a unique $v\in
C^{k+2,\alpha}_\delta$ satisfying
$$
 F_{\mathring g+h}(v)=0
 \;, \quad v>-1\;,\quad
 \|v\|_{C^{k+2,\alpha}_\delta}\le C\|h\|_{C^{k+2,\alpha}_\delta}
 \;,
$$
so that the tensor field $u^{\frac{4}{n-2}}(\mathring g+h)$ defines
a Riemannian metric with constant scalar curvature $-n(n-1)$. The
map $h\mapsto v$ is smooth near zero.
\end{proposition}

Given a Riemannian metric $\mathring g$ as in the statement of
Proposition~\ref{scalconf}, we will use that Proposition to
construct metrics which are arbitrarily close to ${\mathring g}$ on
compact sets, and which are conformal to the background $b$ near
infinity, as follows. Let $0<\delta<\sigma$, $\delta< n$, and let
$g_a$ be the metric interpolating between ${\mathring g}$ and $b$ on
the annulus $A_{a,4a}$, as in \eq{glue0}. Then $g_a={\mathring
g}+h_a$, where $h_a\in C^{k+2,\alpha}_{\sigma} \subset
C^{k+2,\alpha}_{\delta }$, with $\|h_a\|_{C^{k+2,\alpha}_{\delta }}$
going to zero  when $a$ does. Proposition \ref{scalconf} shows that
for all $a$ small enough there exists $v_a\in C^{k+2,\alpha}_{\delta
}$ satisfying $v_a>-1$ such that the metric
\bel{hmeta}
 \hat g_a:= (1+v_a)^{\frac 4 {n-2}}g_a
\ee
has  scalar curvature $-n(n-1)$,   and $v_a$ goes to zero in
$C^{k+2,\alpha}_\delta$ when $a$ does. In particular, $v_a$ goes to
zero with $(k+2,\alpha)$ derivatives uniformly on any compact subset
of $M$.

The metric $\hat g_a$ is conformal to $b$ near the conformal
boundary at infinity. If we assume that $\rho^2 b$ is smooth up
to boundary at $\partial M$, then the conformal factor $u$ is
polyhomogeneous at the conformal boundary~\cite{AndChDiss,ACF}
. Further~\cite{ACF}, the asymptotic expansion of $u_a=1+v_a$
is identical to that of the background metric $b$ up to terms
$O(\rho^n)$.  This implies that $u_a$ is in fact smooth up to
boundary and, for small $\rho$,
\bel{ddec}
 |\hat g_a-b|_b=O(\rho^n)
 \;.
\ee

If $b$ has the form \eq{bmetinr},  {with $\hat b$ -- Einstein}, and
if $\sigma >n/2$, then the energy-momentum vector $p_{(\mu)}$ of
${\mathring g}$ is well defined. We can then choose $\delta>n/2$, in
which case it immediately follows from the definition of $p_{(\mu)}$
and from \eq{eq:3.2}-\eq{eq:3.5} that the energy-momentum vector of
$(1+v_a)^{\frac{4}{n-2}}g_a$ tends to that of ${\mathring g}$ as $a$
goes to zero.

As we have seen, the construction can be done rather generally,
resulting in a small conformal deformation of the metric on
compact sets. It turns out that the deformation can be
localized to the asymptotic region if one supposes, moreover,
that ${\mathring g}$ is \emph{not} static in the asymptotic
region; by this we mean that $P^*_{\mathring g}$ has no kernel
on $M_\epsilon$ for all $\epsilon$ small enough. Then the
deformation can be localized to the exterior region, in the
sense that for any $\epsilon>0$ we can find a constant scalar
curvature metric $\tilde g_\epsilon$ which coincides with
$\mathring g$ on $M\setminus M_\epsilon$, with $M_\epsilon$ as
in \eq{Mdeltadef}, and which is conformal to $b$ near the
conformal boundary. The construction goes as follows: By the
arguments in~\cite{ChBeignokids} there exists a sequence of
annuli $A_{a_i,4a_i}$ on which $P^*_{\mathring g} $ has no
kernel. Choose $a_{i_0}<\epsilon/4$,  for all $a<a_{i_o}$ small
enough let $\hat g_a$ be as in \eq{hmeta}, then $\hat g_a$
restricted to $A_{a_{i_0},4a_{i_0}}$ approaches zero in
$C^{k+2,\alpha}$. By the gluing results
of~\cite{ChDelay,ChDelayHilbert}, for $k > \lfloor
\frac{n}{2}\rfloor + 2$ and for $a$ small enough $\hat g_a$ can
be deformed within $A_{a_{i_0},4a_{i_0}}$ to a metric  $\tilde
g_{\epsilon}$ with constant scalar curvature which coincides
with ${\mathring g}$ on $M\setminus M_{4a_{i_0}}\subset
M\setminus M_\epsilon$, and which coincides with $\hat g_a$ on
$M_{a_{i_0}}$, hence is conformal to $b$ near the conformal
boundary at infinity. In particular $\tilde g_{\epsilon}$
approaches $b$ as $O(\rho^n)$ by \eq{ddec}.

Summarizing, we have proved the following result, somewhat
reminiscent of~\cite[Proposition~4.1]{SchoenCatini}:

\begin{proposition}
 \label{Pdef}
Let $\dim M=n\ge 3$, $C,\sigma>0$, $\ell \in \N$, $\ell \ge 2$,
$\alpha \in (0,1)$ and suppose that ${\mathring g}=b+\zh $ is a
Riemannian metric with scalar curvature $-n(n-1)$ with
$\rho^2b\in C^{l,\alpha}(\bM)$ and $\zh  \in
C^{l,\alpha}_\sigma$. Then:

\begin{enumerate}
 \item For all $\epsilon>0$ there exists a metric
     $\mathring g_\epsilon$ with scalar curvature
     $-n(n-1)$,  conformal to $\mathring g$ away from
     $M_\epsilon$, and  conformal to $b$ near the conformal
     boundary.

\item \label{Pp2} Furthermore $\mathring g_\epsilon$ converges to ${\mathring g}$ in
$C^{\ell ,\alpha}(\mcU)$ topology
on any relatively compact open subset $\mcU$ of $M$.

\item
If $\rho^2 b$ is sufficiently differentiable at the conformal
boundary (e.g., smooth), then the metrics $\mathring g_\epsilon$
approach $b$ as $O(\rho^n)$ for small $\rho$.

\item
If $b$ is of the form \eq{bmetinr} with $\hat b$ Einstein, and if  $\sigma>n/2$, then the energy-momentum  of $\mathring g_\epsilon$ approaches that
of ${\mathring g}$ as $\epsilon$ tends to zero.

 \item
If  $  \ell>\lfloor \frac{n}{2}\rfloor + 4$ and if there exists
$\epsilon_0>0$ such that  $P_{{\mathring g}}^*$ has no kernel on
$A_{\epsilon_0/4,\epsilon_0}$, then $\mathring g_\epsilon$ can be
chosen to coincide with ${\mathring g}$ away from $M_{\epsilon_0}$,
but then the convergence of point (2) to ${\mathring g}$ is  in
$C^{\ell-2 ,\alpha}(\mcU)$ topology only.
 \end{enumerate}
\end{proposition}

\appendix

\section{The asymptotics of $P^*$}
\label{ASc}

\subsection{Conformally compact metrics}

In this section, we study the behaviour  of the operator
$P^*_g$, when rescaled from $A_{\delta,4\delta}$ to  $A_{1,4}$,
with $\delta$ tending to zero, for  conformally compact metric,
asymptotically hyperbolic in the sense of~\cite{mazzeo:unique}.
We consider on $M_{4\delta}$ a metric of the form
\bel{gconfcomp}
 g=\rho^{-2}(d\rho^2+\hat{g}(\rho))=:\rho^{-2}\overline{g}.
 \ee
This metric is conformally compact and
$|d\rho|_{\overline{g}}=1$ at infinity, so
$$
\Ric(g)=-(n-1)g+O(\rho),
$$
where $O(\rho)$ is a symmetric covariant two tensor with $g$-norm
of order $O(\rho)$ (equivalently $\overline{g}$-norm of order
$O(\rho^{-1})$).

We  study the metric   on $A_{\delta,4\delta}$, of course this
calculation is valid for $g=b$ with $b$ as is \eq{methypxx} or \eq{bmetinr2}. This
can be pulled-back to $A=A_{1,4}$ using the change of variable
$\rho=\delta z$ to
$$
\delg =z^{-2}(dz^2+\delta^{-2}\;^\delta\hat{g}(z)),
$$
where $\;^\delta\hat{g}(z)=\hat{g}(\delta z)$. The determinant
reads
$$
\det (\delg)=z^{-2n}\delta^{-2(n-1)}\det(\;^\delta \hat{g})
 \;.
$$
The non-trivial Christoffel symbols of $\delg $ are
$$
\;^\delta\Gamma^z_{zz}=-z^{-1},
$$
$$
\;^\delta\Gamma^z_{AB}=-\frac{z^2}{2}(-2z^{-3}\delta^{-2}\;^\delta\hat{g}_{AB}
+z^{-2}\delta^{-2}\partial_z\;^\delta\hat{g}_{AB}),
$$
$$
\;^\delta\Gamma^C_{Az}=\;^\delta\Gamma^C_{zA}=\frac{1}{2}(-2z^{-1}\delta^C_A
+\;^\delta\hat{g}^{BC}\partial_z\;^\delta\hat{g}_{AB}),
$$
$$
\;^\delta\Gamma^C_{AB}=\Gamma^{C}_{AB}(\;^\delta\hat{g})=:\;^\delta\hat{\Gamma}^{C}_{AB}.
$$
We note that
$\partial_z\;^\delta\hat{g}_{AB}(z)=\delta\partial_\rho\hat{g}_{AB}(\delta
z)=O(\delta).$ The Hessian of a function $u$ takes the form
$$
\;^\delta\nabla_z\partial_zu=\partial_z^2u+z^{-1}\partial_zu, $$
$$
\;^\delta\nabla_z\partial_Au=\partial_z\partial_Au+(z^{-1}\delta^C_A
+O(\delta)^C_A)\partial_Cu,
$$
$$
\;^\delta\nabla_A\partial_Bu=\;^\delta\hat{\nabla}_A\partial_Bu
-(z^{-1}\delta^{-2}\;^\delta\hat{g}_{AB}
+O(\delta^{-1})_{AB})\partial_zu,
$$
thus
$$
\;^\delta\nabla^k\partial_ku=z^2\partial_z^2u-[(n-2)z+O(\delta)]\partial_zu
+z^2\delta^2\;^\delta\hat{\nabla}^A\partial_Au.
$$
This gives
%
$$
(P_\delg ^*u)_{zz}=[(n-1)z^{-1}+O(\delta)]\partial_z u-u\Ric(\delg
)_{zz}-\delta^2\;^\delta\hat{\nabla}^A\partial_Au,
$$
$$
(P_\delg ^*u)_{zA}=\partial_z\partial_Au+(z^{-1}\delta^C_A
+O(\delta)^C_A)\partial_Cu-u\Ric(\delg)_{zA},
$$
\begin{eqnarray*}
(P_\delg
^*u)_{AB}&=&\;^\delta\hat{\nabla}_A\partial_Bu-\;^\delta\hat{\nabla}^C\partial_Cu\;^\delta\hat{g}_{AB}
-\delta^{-2}\partial_z^2u\;^\delta\hat{g}_{AB}\\
&&\;\;
+[(n-3)z^{-1}\delta^{-2}\;^\delta\hat{g}_{AB}+O(\delta^{-1})_{AB}]\partial_zu-u\Ric({g}_\delta)_{AB}.
\end{eqnarray*}

Now, recall that $\Ric(g)=-(n-1)g+\rho^{-1}T$, where $T$ is
$\overline{g}$ bounded. As
$\overline{g}=d\rho^2+\hat{g}(\rho)=\delta^2dz^2+\;^\delta\hat{g}(z)$,
we have that $T_{zz}=O(\delta^2)$, $T_{zA}=O(\delta)$ and
$T_{AB}=O(1)$, thus the coordinate components of $P_\delg ^*u$ are
$$
(P_\delg ^*u)_{zz}=[(n-1)z^{-1}+O(\delta)]\partial_z
u+[(n-1)z^{-2}+O(\delta)]u-\delta^2\;^\delta\hat{\nabla}^A\partial_Au,
$$
$$
(P_\delg ^*u)_{zA}=\partial_z\partial_Au+(z^{-1}\delta^C_A
+O(\delta)^C_A)\partial_Cu-uO(1)_{zA},
$$
\begin{eqnarray*}
(P_\delg
^*u)_{AB}&=&\;^\delta\hat{\nabla}_A\partial_Bu-\;^\delta\hat{\nabla}^C\partial_Cu\;^\delta\hat{g}_{AB}
-\delta^{-2}\partial_z^2u\;^\delta\hat{g}_{AB}\\
&&\;\;
+[(n-3)z^{-1}\delta^{-2}\;^\delta\hat{g}_{AB}+O(\delta^{-1})_{AB}]\partial_zu\\
&&\;\;+u(n-1)(z^{-2}\delta^{-2}\;^\delta\hat{g}_{AB}+O(\delta^{-1})_{AB}).
\end{eqnarray*}

\subsection{The $(C,k,\sigma)$-asymptotically hyperbolic case}

In this section we compare the behaviour of the operator
$P^*_g$ with that of $P^*_b$, when rescaled from
$A_{\delta,4\delta}$ to $A_{1,4}$, for
$(C,k,\sigma)$-asymptotically hyperbolic metrics of the form
\eq{gconfcomp}. We also give an explicit formula for $P_b^*$
and its kernel for metrics of the form \eq{methypxx}.

If $k\in\N$, $\sigma>0$, and $g$ is
$(C,k,\sigma)$-asymptotically hyperbolic with $b$ of the form
\eq{methypxx}, we have
$$
\Ric(g)=\Ric(b)+O(\rho^\sigma)=-(n-1)b+O(\rho^\sigma)=-(n-1)g+O(\rho^\sigma),
$$
where $O(\rho^\sigma)$ is a symmetric covariant two tensor with
$g$-norm (or $b$-norm) of order $O(\rho^\sigma)$ (equivalently
$\overline{g}$-norm of order $O(\rho^{\sigma-2}))$ .

 First, we have
$\hat{g}(\rho)-\hat{b}(\rho)=O(\rho^\sigma)$ and
$\partial_\rho[\hat{g}-\hat{b}](\rho)=O(\rho^{\sigma-1})$, so that
$$\;^\delta\hat{g}(z)-\;^\delta\hat{b}(z)=O(\delta^\sigma)
 \;,
$$
$$\partial_z[\;^\delta\hat{g}-\;^\delta\hat{b}\;](z)=\delta
\;O(\delta^{\sigma-1})=O(\delta^\sigma).$$ The non-trivial
Christoffel symbols of $\delg $ are
$$
\;^\delta\Gamma^z_{zz}=-z^{-1}=\;^\delta\Gamma^z_{zz}(b_\delta),
$$
$$
\;^\delta\Gamma^z_{AB}=\;^\delta\Gamma^z_{AB}(b_\delta)+O(\delta^{\sigma-2})_{AB},
$$
$$
\;^\delta\Gamma^C_{Az}=\;^\delta\Gamma^C_{zA}=\;^\delta\Gamma^C_{Az}(b_\delta)+O(\delta^\sigma)^C_{A},
$$
$$
\;^\delta\Gamma^C_{AB}=\Gamma^{C}_{AB}(\;^\delta\hat{g})=\;^\delta\hat{\Gamma}^{C}_{AB}(^\delta\hat{b})
+O(\delta^\sigma)^C_{AB}.
$$
Let $\nu$ be a one-form on $A$. To make things clear, let
$\;^\delta\widetilde{\nabla}$ denote the covariant derivative
operator of the metric $\bdelta $ and $\;^\delta{\nabla}$ the
one for $g_\delta$, we have
$$
\;^\delta\nabla_z\nu_z=\partial_z\nu_z+z^{-1}\nu_z, $$
$$
\;^\delta\nabla_z\nu_A=\;^\delta\widetilde{\nabla}_z\nu_A
+O(\delta^\sigma)^C_A\nu_C,
$$
$$
\;^\delta\nabla_A\nu_z=
\;^\delta\widetilde{\nabla}_A\nu_z+O(\delta^\sigma)^C_A\nu_C,
$$
$$
\;^\delta\nabla_A\nu_B=\;^\delta\widetilde{\nabla}_A\nu_B
+O(\delta^{\sigma-2})_{AB}\nu_z+O(\delta^{\sigma})^C_{AB}\nu_C,
$$
where the error terms are measured with any fixed metric on the
compact set $\overline {A}$, e.g. $dz^2+\hat b(0)$.

If $\nu=du$, then
$$
\;^\delta\nabla_z\partial_zu=\partial_z^2u+z^{-1}\partial_zu=\;^\delta\widetilde{\nabla}_z\partial_zu,
$$
$$
\;^\delta\nabla_z\partial_Au=
\;^\delta\widetilde{\nabla}_z\partial_Au+O(\delta^\sigma)^C_A\partial_Cu,
$$
$$
\;^\delta\nabla_A\partial_Bu=
\;^\delta\widetilde{\nabla}_A\partial_Bu+O(\delta^{\sigma-2})_{AB}\partial_zu
+O(\delta^{\sigma})^C_{AB}\partial_Cu,
$$
thus
$$
\;^\delta\nabla^k\partial_ku=\;^\delta\widetilde{\nabla}^k\partial_ku+O(\delta^\sigma)\partial_zu
+O(\delta^{\sigma+2})^{AB}\;^\delta\hat{\widetilde{\nabla}}_B\partial_Au+O(\delta^{\sigma+2})^C\partial_Cu.
$$
We obtain for the components of $P_\delg ^*u$:
$$
(P_\delg ^*u)_{zz}=(P_{\bdelta }
^*u)_{zz}+O(\delta^\sigma)\partial_zu +O(\delta^\sigma)u
+O(\delta^{\sigma+2})^{AB}\;^\delta\hat{\widetilde{\nabla}}_B\partial_Au
+O(\delta^{\sigma+2})^C\partial_Cu,
$$
$$
(P_\delg ^*u)_{zA}=(P_{\bdelta }
^*u)_{zA}+O(\delta^\sigma)^C_A\partial_Cu+O(\delta^{\sigma-1})_Au,
$$
$$
(P_\delg ^*u)_{AB}=(P_{\bdelta }
^*u)_{AB}+O(\delta^{\sigma-2})_{AB}\partial_zu
+O(\delta^{\sigma})^C_B\;^\delta\hat{\widetilde{\nabla}}_C\partial_Au+O(\delta^{\sigma})^C_{AB}\partial_Cu+
O(\delta^{\sigma-2})_{AB}u
 \;.
$$

Next, we compute the explicit expression of $P_{\bdelta } ^*$
for a metric of the form \eq{methypxx}. In that case we have
$$
^\delta\hat{b}(z)=\left[1-k\left(\frac{\delta
z}{2}\right)^2\right]^2\hat{b},
$$
then
$$
\partial_z({}^\delta\hat{b})(z)=-k\delta^2z\left[1-k\left(\frac{\delta
z}{2}\right)^2\right]\hat{b}.
$$

The non-trivial Christoffel symbols of  ${b}_\delta$ are
$$
\;^\delta\Gamma^z_{zz}=-z^{-1},
$$
$$
\;^\delta\Gamma^z_{AB}=\delta^{-2}z^{-1}\left[1-k\left(\frac{\delta
z}{2}\right)^2\right]\left[1+k\left(\frac{\delta
z}{2}\right)^2\right]\hat{b}_{AB},
$$
$$
\;^\delta\Gamma^C_{Az}=-z^{-1}\left[1-k\left(\frac{\delta
z}{2}\right)^2\right]^{-1}\left[1+k\left(\frac{\delta
z}{2}\right)^2\right]\delta^C_A,
$$
$$
\;^\delta\Gamma^C_{AB}={\Gamma}^{C}_{AB}(\hat{b}).
$$
We thus obtain for the components of the Hessian of $u$:
$$
\;^\delta\widetilde{\nabla}_z\partial_zu=\partial_z^2u+z^{-1}\partial_zu, $$
$$
\;^\delta\widetilde{\nabla}_z\partial_Au=\partial_z\partial_Au+
z^{-1}\left[1-k\left(\frac{\delta
z}{2}\right)^2\right]^{-1}\left[1+k\left(\frac{\delta
z}{2}\right)^2\right]\partial_Au,
$$
$$
\;^\delta\widetilde{\nabla}_A\partial_Bu=\hat{\nabla}_A\partial_Bu
-\delta^{-2}z^{-1}\left[1-k\left(\frac{\delta
z}{2}\right)^2\right]\left[1+k\left(\frac{\delta
z}{2}\right)^2\right]\hat{b}_{AB}\;\partial_zu,
$$
thus
$$
\begin{array}{lll}
\;^\delta\widetilde{\nabla}^k\partial_ku&=&\displaystyle
z^2\partial_z^2u+\left\{1-(n-1)\left[1-k\left(\frac{\delta
z}{2}\right)^2\right]^{-1}\left[1+k\left(\frac{\delta
z}{2}\right)^2\right]\right\}z\partial_zu\\
&&\displaystyle\hspace{4cm}+z^2\delta^2\left[1-k\left(\frac{\delta
z}{2}\right)^2\right]^{-2}\hat{\nabla}^A\partial_Au.
\end{array}
$$
One checks that $\Ric(b)=-(n-1)b$, which implies
 $$\Ric(\bdelta)=-(n-1)\bdelta .$$
We obtain for the components of $P_{\bdelta } ^*u$:
$$
\begin{array}{lll}
 \displaystyle
(P_{\bdelta } ^*u)_{zz}&=&
 \displaystyle(n-1)\Bigg\{\left[1-k\left(\frac{\delta
z}{2}\right)^2\right]^{-1}\left[1+k\left(\frac{\delta
z}{2}\right)^2\right]z^{-1}\partial_zu+z^{-2}u\\
 \displaystyle
&&
 \displaystyle\;\;\;\;\;\;\;\;\;\;\;\;\;\;\;-\frac{1}{n-1}\delta^2\left[1-k\left(\frac{\delta
z}{2}\right)^2\right]^{-2}\hat{\nabla}^A\partial_Au\Bigg\},
\end{array}
$$
$$
(P_{\bdelta } ^*u)_{zA}=\partial_z\partial_Au
+z^{-1}\left[1-k\left(\frac{\delta
z}{2}\right)^2\right]^{-1}\left[1+k\left(\frac{\delta
z}{2}\right)^2\right]\partial_Au,
$$
\begin{eqnarray*}
(P_{\bdelta }
^*u)_{AB}&=&\displaystyle\hat{\nabla}_A\partial_Bu-\hat{\nabla}^C\partial_Cu\;\hat{b}_{AB}
-\delta^{-2}z^{-2}\left[1-k\left(\frac{\delta
z}{2}\right)^2\right]^{2}\Bigg\{z^2\partial^2_zu\\
&&\;\;+z\left(1-(n-2)\frac{1+k\left(\frac{\delta
z}{2}\right)^2}{1-k\left(\frac{\delta
z}{2}\right)^2}\right)\partial_zu-(n-1)u \Bigg\} \hat{b}_{AB}.
\end{eqnarray*}

Now, a function $u$ is in  the kernel of $P_{\bdelta } ^*$ if
and only if
$$
\;^\delta\nabla\partial u=u\;b_\delta.
$$
One checks that
$$
u_0=z^{-1}\left[1+k\left(\frac{\delta z}{2}\right)^2\right],
$$
is indeed in  this kernel. Further, if $v$ is a non trivial solution
of the Obata-type equation
$$
 \hat{\nabla}_A\partial_Bv=-kv\;\hat{b}_{AB}
$$
on the boundary at infinity, then the function
$$
 u=v
 z^{-1}\left[1-k\left(\frac{\delta z}{2}\right)^2\right]
$$
satisfies $(P_{\bdelta } ^*u)_{zz}=(P_{\bdelta } ^*u)_{zA}=0$,
with ($(P_{\bdelta } ^*u)_{AB}=0$ if and only if $k=1$).
Finally, it is an easy exercise to show that these functions
generate the kernel of $P_{\bdelta } ^*$.  Here one can use the
well known fact that the kernel of $P^*$ has dimension at most
$n+1$ (see, e.g.,~\cite{Corvino}).

\section{Proof of Lemma~\ref{Lhdes2}}
\label{A1}

Throughout this appendix we write $A_\delta$ for
$A_{\delta,4\delta}$ and $A$ for $A_{1,4}$; we hope that the
clash of notation with the $A$--spaces occasionally used
elsewhere will not confuse the reader. We start by scaling
$A_\delta$ to $A=(1,4)\times
\partial M$. Recall that the weight function $\phi=y^2$ on $A$
relevant for the calculation at hand equals $(z-1)^2(z-4)^2/9$,
where $z$ runs along the $(1,4)$ factor of $A$. The argument
that follows actually applies to any non-negative function
$\phi=\phi(z)$ which vanishes precisely at the boundary of $A$
and satisfies:
$$
 \phi(1)=\phi(4)=\phi'(1)= \phi'(4)=0\;,\quad \phi''(1)>0\;,
 \phi''(4)> 0
 \;.
$$

The idea of the proof is to cover  $(1,4)$ by intervals $I_i$,
with sizes chosen so that on each interval the ratio $\sup
\phi/\inf \phi$  is bounded independently of $i$. Furthermore,
the size of each interval should be of the order of the value
of $\phi$ on the interval, to ensure good scaling properties.
We then use interior elliptic estimates after a cube
decomposition of $I_i \times
\partial M$; this requires a second family of  thickened intervals
$\hat I_i$, with properties similar to the ones satisfied by
the $I_i$'s. Summing over the cubes provides the desired
estimate, after having ensured that the $\hat I_i$'s do not
overlap too much. We note that the scalings in $z$ and
$\theta^A$ are different; the former is tailored to account for
the degeneracy in the ``radial" $z$-direction,  measured by
$\phi$, and the latter accounting for the
$\phi\delta$-dependent degeneracy in the ``angular"
$\theta^A$-direction.

So we divide $(1,4)$ into intervals
\bel{wantik}
 I_k\subset \hat I_k\subset (1,4)\;,\quad \cup_k I_k = (1,4)\;,
\ee
as follows: There exists $1<z_1<5/2$ such that $\phi:[1,z_1]\to
\R^+$ is strictly increasing. Choose $a>0$ small enough so
that
$$
z-2a
 \phi(z)>1 \mbox{\ on $(1,z_1]$ and \ } z_1+a(\phi(z_1))\le 4
  \;.
$$
Define $z_i$ by induction using
$$
 z_{i+1}= z_i - a\phi(z_i)
 \;,
$$
thus $1<z_{i+1}<z_i$, and $\lim_{i\to \infty} z_i =1$. For any
function  $f\in L^1(A)$ we thus have
\bel{firint}
 \int_{[1,z_1]\times \partial M}f = \sum_i \int_{[z_{i+1},z_i]\times \partial M} f
 \;.
\ee
%
We want to show that  there exists a constant $C$ such that for all
$a$ small enough and for all positive integrable functions $f$ we
have
\bel{firint2}
  \sum_i \int_{[z_{i}-2a\phi(z_{i}),z_i+a\phi(z_{i})]\times \partial M}
  f \le C  \int_{[1,4]\times \partial M}f
 \;.
\ee
%
In order to do that we need to count how many of the intervals
$[z_{i}-2a\phi(z_{i}),z_i+a\phi(z_{i})]$ overlap. Letting $b:=
a \phi''(1)/2$, one easily finds
\beaa
 z_i - 2a\phi(z_i)-1 &= & (z_i-1)\Big(1-2b(z_i-1)\Big)+ O\Big((z_i-1)^3\Big)\;,
 \\\
 z_{i+k} -1 &= & (z_i-1)\Big(1-kb(z_i-1)\Big)+ O\Big((z_i-1)^3\Big)\;,
 \\
 z_{i+k} + a \phi(z_{i+k})-1 &= &
(z_i-1)\Big(1-(k-1)b(z_i-1)\Big)+ O\Big((z_i-1)^3\Big)\;, \eeaa
where the error terms in the second and third equation depend
upon $k$. Choosing $k=4$, it follows that
$$
z_{i+k} + a \phi(z_{i+k}) < z_i -2 a\phi(z_i)
$$
for all $i$ large enough. So for $i$ large enough
$[z_{i}-2a\phi(z_{i}),z_i+a\phi(z_{i})]$ will intersect at most
six such other intervals, and \eq{firint2}  with  a constant
$C\geq6$ follows.

An obvious modification of the above construction, decreasing $a$
if necessary, will lead to a sequence $\hpz _k\to 4$ satisfying
$$
 5/2 < \hpz _1 \le \hpz _i = \hpz _{i+1} - a \phi(\hpz _{i+1})<\hpz _{i+1} < 4
 \;,
$$
with, for $f\in L^1(A)$,
\bel{firint3}
 \int_{[\hpz _1,4]\times \partial M}f = \sum_i \int_{[\hpz _{i},\hpz _{i+1}]\times \partial M} f
 \;,
\ee
and if moreover $f$ is  positive then
\bel{firint4}
  \sum_i \int_{[\hpz _{i}-2a\phi(\hpz _{i}),\hpz _{i}+a\phi(\hpz _{i})]\times \partial M}
  f \le C  \int_{[1,4]\times \partial M}f
 \;.
\ee
%

Letting $\{I_k\}_{k\in \N}$ be the collection, without
repetitions, of the intervals
$$
 \{\underbrace{[z_1,\hpz _1]}_{=:I_1},[z_{i+1},z_i],[\hpz _j,\hpz_{j+1}]\}_{i,j\in \N^*}
 \;,
$$
and letting $\{\hat I_k\}_{k\in \N}$ be the collection, without
repetitions, of the intervals
$$
 \{\underbrace{[z_1-a\phi(\hpz_1),\hpz _1+a\phi(\hpz _1)]
 }_{=:\hat I_1},[z_{i}- 2a \phi(z_{i}),z_i+a \phi(z_{i})],[\hpz _j-2a \phi(\hpz _{j}),
 \hpz_{j}+a \phi(\hpz _{j})]\}_{i,j\in \N^*}
 \;,
$$
we obtain \eq{wantik} together with
\beal{intik}
 &
 \displaystyle
 \int _{[1,4]\times \partial M} f  = \sum_k \int_{I_k\times \partial M}f
 \;,
 &
 \\
 &
 \displaystyle
 C^{-1} \int _{[1,4]\times \partial M} f  \le  \sum_k \int_{\hat I_k\times \partial
 M}f \le C \int _{[1,4]\times \partial M} f
 \;,
 &
\eeal{intik2}
for positive $f\in L^1(A)$. We set
$$
 \tilde z_k = \sup I_k\;.
$$

The above construction provides a $\delta$-independent decomposition
of $A$ into stripes $I_k\times \partial M$, the size of which in the
$z$-direction is comparable to $\phi(z)$ for any $(z,v)\in I_k\times
\partial M$; similarly the sizes of $\hat I_k\times \partial M$ are comparable to $\phi(z)$ for any $(z,v)\in \hat I_k\times
\partial M$.
Mapping $A$ to $A_\delta$ provides an associated decomposition of
$A_\delta$ into stripes $I_k^\delta\times \partial M$  with sizes
uniformly comparable to $\phi(\rho/\delta)$ for any $(\rho,v)\in
I_k^\delta\times
\partial M$; similarly for $\hat I_k^\delta\times \partial M$.

We continue with a $\delta$--dependent, and stripe dependent,
cube decomposition of $\partial M$, as follows: Let
$\{(\mcO_i,\psi_i)\}_{i=1,\ldots,N}$  be a covering of $M$ by
coordinate charts with each coordinate system $\psi_i^{-1}$
mapping $\mcO_i$ smoothly and diffeomorphically to a
neighbourhood of $[0,1]^{n-1}$; the local coordinates on
$[0,1]^{n-1}$ will be denoted by $\theta^A$. We further assume
that $\cup \psi_i([0,1]^{n-1})$ covers $\partial M$ as well.
Let $\varphi_i$ be an associated decomposition of unity, thus
$\sum_i \varphi_i=1$. Setting $f_i=(\varphi_i f)\circ\psi_i $,
for any integrable function $f$ we have
$$
\int_{[1,4]\times \partial M}f = \sum_{i=1}^N \int_{[1,4]\times
[0,1]^{n-1}} f_i
 \;.
$$

Given $\delta$ satisfying $0<\delta<1/\sup_{[1,4]} \phi$ and
given an interval $\Ik$ define $m=m(k,\delta)\in \N$ by the
inequality
\bel{mineq}
 \frac 1 {m+1} \le \phi(\tilde z_k) \delta < \frac 1 m
 \;.
\ee

Let  $\{K_j\}$ be the collection of closed $(n-1)$-cubes, with
pairwise disjoint interiors, and with edges of size $1/m$,
covering $[0,1]^{n-1}$. For any $K_j$ let $\hat K_j$ be the union
of those cubes $K_i$ which have non-empty intersection with $K_j$.
Note that there exists a number $\hat N(n)$ such that $\hat K_j$
is consists of at most $\hat N(n)$ cubes $K_i$. It follows that
for any integrable function $f_i$ we have
$$
 \int _{[1,4]\times
[0,1]^{n-1}} f_i = \sum _{k}\int_{[1,4]\times {K_k}} f_i
 \;,
$$
and if $f_i\ge 0$ then
$$
 \int _{[1,4]\times
[0,1]^{n-1}} f_i \le  \sum _{k}\int_{[1,4]\times {\hat K_k}} f_i
 \le
 \hat N(n)
  \int _{[1,4]\times
 [0,1]^{n-1}} f_i \;.
$$

 We are ready now to pass to the heart of our argument. Let
$\{\mcU_\ell\}_{\ell\in \N}$ be the collection, without repetitions,
of the sets
$$
 \{ \Ik \times \psi_i( K_j)\}_{k\in \N,\ i=1,\ldots,N,\ j=0,\ldots,m^{n-1}}
 \;.
$$
Similarly let $\{\hmcU_\ell\}_{\ell \in \N}$ be the collection,
without repetitions, of the sets
$$
 \{\hat \Ik \times \psi_i(\hat K_j)\}_{k\in \N,\ i=1,\ldots,N,\ j=0,\ldots,m^{n-1}}
 \;.
$$
{}From  what has been said we have, for any positive integrable
function $f$,
\beaa
 &
 \displaystyle
 \int_{[1,4]\times \partial M} f \le  \sum _\ell \int_{\mcU_\ell}
f
 \le
 N \int_{[1,4]\times \partial M} f
 \;,
 &
 \\
 &
 \displaystyle
 \int_{[1,4]\times \partial M} f \le  \sum _\ell \int_{\hmcU_\ell} f \le N \hat N(n) \int_{[1,4]\times \partial M} f
 \;.
 &
\eeaa
If $\mcU_\ell= \Ik \times \psi_i(K_j)$ set
$\phi_\ell=\phi(\tilde z_k)$. Scale the local coordinates
$(z,\theta^A)$ in $\hmcU_\ell$ as
$$
(z,\theta^A)\mapsto (z/\phi_\ell ,m \theta^A)\;.
$$
Up to translations, this maps all $\mcU_\ell\subset \hmcU_\ell$'s to
 fixed cubes
$$
\mcU_\ell \longrightarrow [0,a]\times [0,1]^n\subset [-a,2a]\times
[-1,2]^n \longleftarrow \hmcU_\ell
 \;,
$$
except for those which correspond to $I_1\times \psi_i(K_j)$, which
are mapped to
$$
\mcU_\ell \longrightarrow [0,a(\hpz_1-z_1)/\phi(\hpz_1)]\times
[0,1]^n\subset [-a,a+a(\hpz_1-z_1)/\phi(\hpz_1)]\times [-1,2]^n
\longleftarrow \hmcU_\ell
 \;,
$$
By construction there exists a constant $C>0$, independent of $i$,
such that we have
$$
  {\sup_{I_i \times \partial M}} \phi\le C {\inf_{I_i \times \partial M}}
  \phi\;, \qquad
  {\sup_{\hat I_i \times \partial M}} \phi\le C {\inf_{\hat I_i \times \partial M}}
  \phi
  \;,
$$
hence the same is true on each $\mcU_\ell$ and $\hmcU_\ell$. Let
$\psi = e^{-s/2y}$; it is shown at the end
of~\cite[Appendix~B]{ChDelay} that one also has
$$
  {\sup_{I_i \times \partial M}} \psi\le C {\inf_{I_i \times \partial M}}
  \psi\;, \qquad
  {\sup_{\hat I_i \times \partial M}} \psi\le C {\inf_{\hat I_i \times \partial M}}
  \psi
$$
(with perhaps a different constant $C$), and again such
$\ell$-independent inequalities hold on the $\mcU_\ell$'s and
$\hmcU_\ell$'s.  At this stage in is important to realize that
$$L_{g_\delta}=B(\phi\partial_z,\phi\delta\partial_{\theta^A}),$$
where $B$ is {\em uniformly  elliptic} of order 4 (see equation
(A.4) in~\cite{ChDelay}) on the relevant cubes. We can also
write
$$g_\delta=z^{-2}(dz^2+\delta^2\hat{g}_{AB}(\delta z)d\theta^Ad\theta^B)=
z^{-2}(dz^2+\hat{g}_{AB}(\delta
z)d(\delta\theta)^Ad(\delta\theta)^B).$$ It then follows from
the usual elliptic interior estimates~\cite[p.~246]{Morrey}
 for the operator $B$ and scaling that (here $g=g_\delta$)
$$
 \sum_{i \le k+4} \int_{\mcU_\ell}\psi^2 \phi^{2i} |\nabla^{(i)}u|_g^2 \le C
\Bigg( \sum_{i \le k} \int_{\hmcU_\ell} \psi^2
\phi^{2i}|\nabla^{(i)}L u|_g^2 +
  \int_{\hmcU_\ell} \psi^2| u|^2 \Bigg)\;,
$$
 where $C$ does not depend upon $\delta$ nor $g$ close to
$b$.
Summing over $\ell$, Lemma~\ref{Lhdes2} follows.
\qed

\bigskip

\noindent{\sc Acknowledgements} We are grateful to two
anonymous referees for suggesting many improvements to the
original version of this paper.

\bibliographystyle{amsplain}
\bibliography{../references/newbiblio,%
../references/reffile,%
../references/bibl,%
../references/erwbiblio,%
../references/hip_bib,%
../references/newbib,%
../references/PDE,%
../references/netbiblio,%
../references/dp-BAMS}
\end{document}